\documentclass[12pt]{article}
\usepackage[utf8x]{inputenc}

\usepackage[comma,authoryear]{natbib}
\usepackage{authblk}
\usepackage{subfig}
\usepackage{afterpage}
\usepackage[ruled,vlined]{algorithm2e}
\usepackage{bm}
\usepackage{amsmath}
\usepackage{graphicx}
\usepackage[letterpaper, portrait, margin=1in]{geometry}
\usepackage{amssymb}
\usepackage{adjustbox}

\usepackage{lineno}

\title{Statistical Evaluation of Spectral Methods for Anomaly Detection in Static Networks}
\author[1]{Tomilayo Komolafe\thanks {tomilayo@vt.edu}}
\author[1,2]{A. Valeria Quevedo\thanks{anavq@vt.edu}}
\author[1]{Srijan Sengupta \thanks{sengupta@vt.edu} }
\author[1]{William H. Woodall \thanks{bwoodall@vt.edu}}
\affil[1]{Virginia Polytechnic Institute and State University, Blacksburg VA}
\affil[2]{Universidad de Piura, Peru}

\begin{document}
\maketitle
\textbf{\abstractname{:}}
The topic of anomaly detection in networks has attracted a lot of attention in recent years, especially with the rise of connected devices and social networks. 
Anomaly detection spans a wide range of applications, from detecting terrorist cells in counter-terrorism efforts to identifying unexpected mutations during ribonucleic acid (RNA) transcription. 
Fittingly, numerous algorithmic techniques for anomaly detection  have been introduced. 
However, to date, little work has been done to evaluate these algorithms from a statistical perspective.
This work is aimed at addressing this gap in the literature by carrying out statistical evaluation of a suite of popular spectral methods for anomaly detection in networks.
Our investigation on statistical properties of these algorithms reveals several important and critical shortcomings that we make methodological improvements to address.
Further, we carry out a performance evaluation of these algorithms using simulated networks and extend the methods from binary to count networks.


\section{Introduction}
A network consists of nodes, which represent individual entities, and relationships between nodes, represented as edges \citep{bader2008snap,woodall2017overview}. Investigators in recent years have demonstrated that many phenomena can be represented as networks \citep{woodall2017overview}. These phenomena can span a multitude of fields such as the power grid \citep{albert2004structural, dahan2017network} where nodes represent power stations and edges represent transmission lines, social networks where nodes represent individuals and interactions between individuals depicted as edges \citep{mazrae2016statistical, savage2014anomaly}, or gene sequencing where the nucleotides that make up DNA and RNA during transcription are represented as network motifs \citep{cer2011introducing, cer2012searching, procter2010visualization, raulf1998analysis}. Networks are therefore capable of visually and mathematically representing applications in a myriad of fields \citep{mall2013kernel,woodall2017overview}. 

For this reason, methods that can be applied to networks to identify  abnormalities in these various applications are of significant importance. This is termed the anomaly detection problem where the primary aim is to identify the nodes that are behaving outside of normal conditions \citep{woodall2017overview, miller2015spectral,sengupta2018anomaly}. For example, it is useful to a practitioner to identify over-burdened power plants in a power grid network \citep{dahan2017network}, for a security agency to identify a clandestine operation such as a terrorist cell in a large social network \citep{mazrae2016statistical, savage2014anomaly}, or to identify a series of abnormal proteins in a gene transcription process \citep{cer2012searching, cer2011introducing, procter2010visualization}. Anomaly detection tools allow practitioners to detect unusual behavior in a wide variety of fields \citep{woodall2017overview,akoglu2010oddball,dahan2017network}.
Typically, anomaly detection techniques focus on defining what conditions constitute a normal network  and discriminating between anomalous nodes and non-anomalous nodes  \citep{woodall2017overview, dahan2017network}. 

Networks can be \textit{static}, where we have a single snapshot of the system, or \textit{dynamic}, where we have network snapshots at several points in time.
Anomalies can have different meanings in these two scenarios
\citep{woodall2017overview, savage2014anomaly, FaloustousDynamic}.
In dynamic networks, an anomaly typically corresponds to a group of nodes behaving in a manner that is significantly different from past behavior.
The general approach for detecting such anomalies is to extract some features of the network (such as centrality measures, degree distribution, etc.), monitor these features over time, and raise a signal  when these observed features cross a specified threshold.
A rich class of anomaly detection techniques have been developed for dynamic networks, e.g., density based techniques \citep{papadimitriou2003loci}, clustering based techniques \citep{wang2012identify}, distribution based techniques \citep{akoglu2015graph, vsaltenis2004outlier}, and scan methods \citep{priebe2005scan}.
On the other hand, the goal of anomaly detection in a static network is to detect a subgraph that is significantly different from the overall network \citep{miller2015spectral,sengupta2018anomaly}. 
Some popular approaches include network analysis at the egonet level \citep{akoglu2015graph,sengupta2018anomaly}, spatial autocorrelation  \citep{Chawla2006}, and modularity maximization \citep{newman2016community,sun2005neighborhood, haveliwala2003topic}. 
In our paper, we restrict our attention to static networks.

The critical factors to consider in an anomaly detection problem are the order of the network, the size of the anomalous subgraph to be detected, and the types of anomalies that are of interest \citep{miller2015spectral,dahan2017network}. For example, a small anomalous subgraph is harder to detect than a large anomalous subgraph in the same network \citep{miller2015spectral}. Also, the type of anomalous subgraphs to detect will significantly affect the efficacy of the proposed method \citep{miller2015spectral}. Anomaly detection techniques that are robust to these critical factors are, therefore, highly sought after by practitioners. 

Recently, investigators \citep{miller2010L1norm,miller2015spectral,MillerSparsePCA} developed a suite of anomaly detection methods for static networks based on spectral properties (i.e., eigenvalues and eigenvectors) that are robust to these critical factors.
In particular, the authors demonstrated the applicability of their methods for detecting different types of anomalous subgraphs that are in some instances smaller than 1\% of the network order.
In \citep{miller2015spectral} three spectral methods were proposed, namely the chi-square algorithm, the $L_1$ norm algorithm, and the Sparse Principal Component Analysis (PCA) algorithm.
Of these, the Sparse PCA method has some significant limitations in its implementation due to its computational complexity and the need to tune parameters. 
This method requires estimating the sparse matrix of an eigenspace which is an NP hard problem \citep{miller2015spectral}. 

Any anomaly detection technique consists of computing a network metric (in statistical terms, the \textit{test statistic}) and comparing its value to a benchmark distribution (in statistical terms, the \textit{null distribution}) which represents the distribution of the metric in absence of an anomaly.
If the value of the metric exceeds a threshold obtained from the benchmark distribution, an anomaly is signaled.
There can be two kinds of errors in this process: false alarms that happen when the value of the metric exceeds the threshold although there is no anomaly, and detection failures when the value of the metric is below the threshold in spite of an anomaly.
A principled statistical evaluation is therefore critically important to systematically study whether the following three related criteria are satisfied for a wide range of scenarios: first, the network metric should closely follow the benchmark distribution when there is no anomaly; second, the probability of false alarms is low and close to target values; and third, the probability of detection failures is low and close to target values.
However, there has been relatively little work in such evaluation of anomaly detection techniques.
In this paper we address this gap in the literature by carrying out a systematic statistical evaluation of the spectral methods proposed in \cite{miller2015spectral} for anomaly detection in static networks.
In the interest of space, we restrict our analysis to the chi-square and $L_1$ algorithms. The Sparse PCA method is not covered due to its implementational issues as reported by \cite{miller2015spectral}.   

In \cite{miller2015spectral}, the only case that is explored for both the chi-square and $L_1$ algorithms is a binary network with 4096 vertices and average degree of twelve.  However, it is possible that the algorithms perform differently when the network order or sparsity of the network changes. Hence, in our paper we evaluate the performance of these algorithms for various network order and average degree combinations to provide practitioners with insights on how the algorithms perform under various conditions. Also, in \cite{miller2015spectral}, there is little discussion on establishing a signaling threshold. This task is also very important if these algorithms are to be implemented, and we address this issue comprehensively in our paper. We also demonstrate the effectiveness of these algorithms when applied to count networks, an area not explored in \cite{miller2015spectral} or by other investigators. Our main contributions are summarized below: 

\begin{itemize}
\item{We evaluate the chi-square algorithm and $L_1$ norm algorithm, and identify critical shortcomings pertaining to their statistical properties as well as implementability.}
\item{We introduce methodological improvements to both algorithms. Specifically we provide more practical and appropriate signaling and detection schemes for both algorithms.}
\item{We extend the algorithms to count networks.}
\end{itemize}

To keep our evaluation fair and consistent with \cite{miller2015spectral}, we consider unlabeled static networks generated from the three models used in \cite{miller2015spectral}.
Additionally, simulations rather than case studies are used in our paper to evaluate the methods. 
With simulations, anomalies can be introduced in a controlled manner and the ability to detect particular types of anomalies can be tested \citep{woodall2017overview,savage2014anomaly,azarnoush2016monitoring}. 

The rest of our paper is organized as follows. In section 2, we describe the mathematical formulations used in defining the spectral properties of the networks. In section 3, we describe the chi-square algorithm and carry out a systematic statistical evaluation including some recommendations and methodological improvements. We evaluate the $L_1$ norm algorithm similarly in section 4. We report the performance comparisons of the algorithms for binary networks in section 5. In section 6, we introduce the application of both algorithms to count networks. 
In section 7 we conclude by discussing key points and outlining future research directions. In addition, there are supplementary materials found in the supplementary document that contain additional figures and tables from our multiple simulation comparisons.

\section{Model Setup}
In this section, we discuss the formulation of the residual matrix that is used in the ensuing algorithms. We also describe the formulation of the three network models  along with the spectral properties of their residual matrices \citep{miller2015spectral}.

\subsection{Mathematical Definitions}
A network $G$ = $(V,E)$ is composed of a set of vertices, $V$, and a set of edges, $E$. A subgraph of such a network is a smaller network $G_s = (V_s, E_s)$ such that all vertices and edges of the subgraph belong to the original network, i.e., $V_s\subset V, E_s\subset E$. 
The number of vertices and edges in a network graph $G$ are denoted by $n$ and $M$ respectively,  i.e., $n = |V|,M = |E|$. 
A network can also be represented as an $n \times n$ adjacency matrix denoted by $A$, where $A_{ij}$ is the number of edges between vertices $i$ and $j$, for $i,j = 1, \ldots, n$.

For binary networks, $A_{ij}$ can be $0$ or $1$ and we assume $A_{ij} \sim Bernoulli(p_{ij})$.
For count networks, $A_{ij}$ can be any non-negative integer and we assume $A_{ij} \sim Poisson(\lambda_{ij})$.
Using the same terminology as \cite{miller2015spectral}, the residual matrix of the observed adjacency matrix $A$ is
\begin{equation} \label{eq_expected}
B = A - E[A],
\end{equation}
where $E[A]$ is the expectation of $A$. The $i^{th}$ and $j^{th}$ elements of the matrix $E[A]$ is $p_{ij}$ for binary networks and $\lambda_{ij}$ for count networks. 

Both methods studied in our paper use the spectral structure of the residual matrix $B$ for anomaly detection.
The spectral structure of a square matrix refers to its eigenvalues and eigenvectors.
Note that the residual matrix $B$ is a square matrix with $n$ rows and $n$ columns, and therefore, it can be factorized as 
\[
B = U \Lambda U',
\]
where $U$ is an $n$-by-$n$ matrix whose $i^{th}$ column is the $i^{th}$ eigenvector of $B$, and $\Lambda$ is the diagonal matrix whose diagonal elements are the corresponding eigenvalues.
This factorization, known as eigendecomposition, is a foundational technique for statistical methods based on dimension reduction.
The spectral structure of networks has been widely used in the network literature for developing analytical methods, some such methodological papers are \cite{MillerSparsePCA,miller2010L1norm,rohe2011spectral,qin2013regularized,mall2013kernel,sengupta2015spectral,lei2015consistency}, and \cite{miller2015spectral}.

\subsection{Network models}
\begin{figure}[!hbt]
	\begin{center}
		\includegraphics[height=0.2 \textheight]{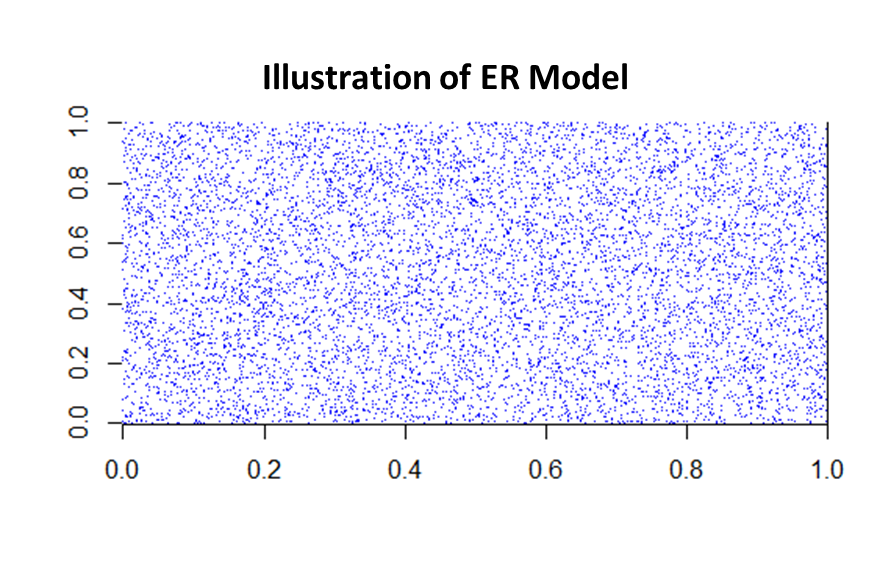}
		\includegraphics[height=0.2 \textheight]{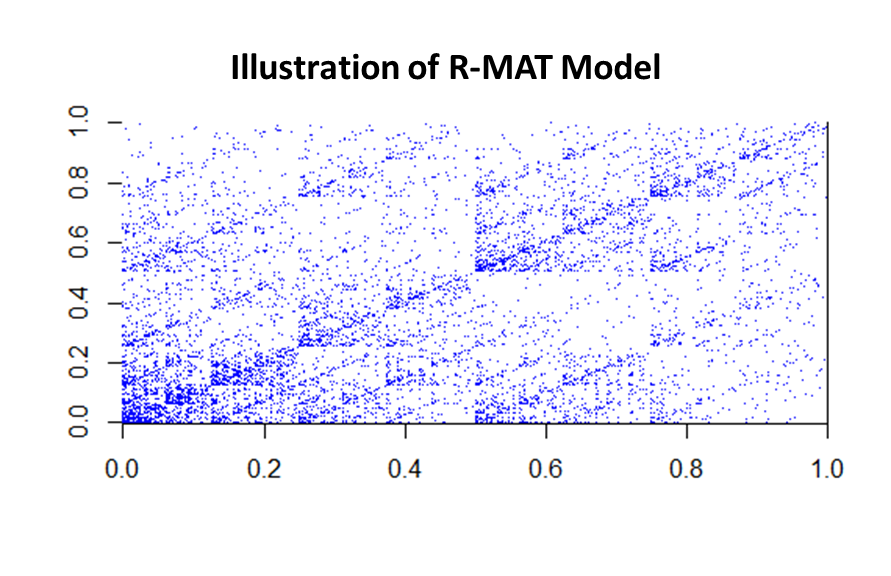} \\
		\includegraphics[height=0.2 \textheight]{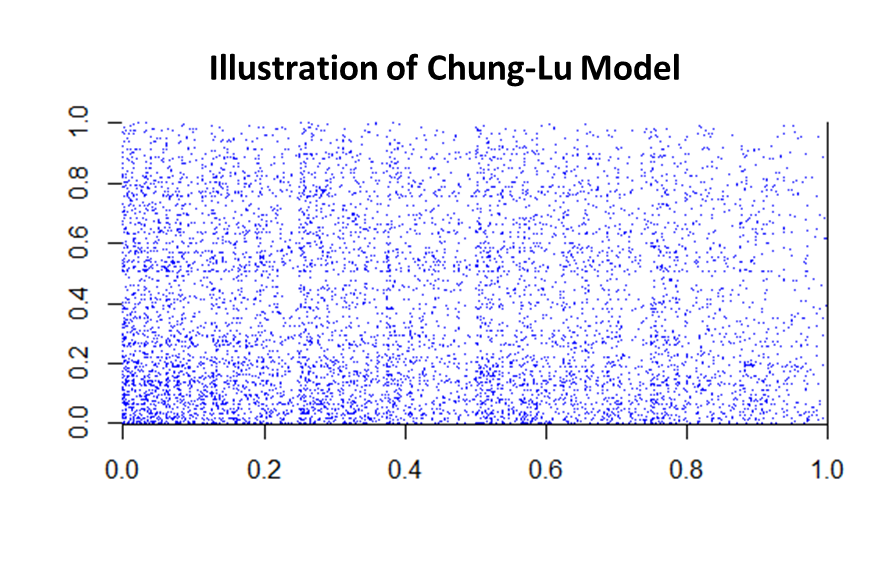}
		\caption{Adjacency matrix illustration with $n = 1024$. Top left: ER model with $p_0 = 0.1$. Top right: R-MAT model with $E \approx 100000$, $a$ = 0.5, $d$ = 0.25, $b$ = 0.125 and $c$ = 0.125. Bottom: Chung-Lu model with $E \approx 100000$ and expected degrees from R-MAT.
		The dots represent 1's, and whitespace corresponds to 0. 
		Note the difference between network structures, which makes it challenging to develop a method that can detect an embedded anomalous subgraph (e.g., a small clique) equally well for the three networks.}
		\label{fig:ER}
	\end{center}
\end{figure}

In \cite{miller2015spectral}, three types of network models with varying complexities are introduced. 
The models are the Erd\"os-R\'enyi model, the R-MAT model, and the Chung-Lu model. 
Figure \ref{fig:ER} is a visualization of adjacency matrices generated from the three models.
Their formulations are described below. 

Erd\"os-R\'enyi (ER) networks are simple networks that are generated given only a single parameter, the background probability $p_{ij} = p_0  \forall  i,j$ or $\lambda_{ij} = \lambda_0 \forall i,j$ \citep{erdos1960evolution, miller2015spectral,chung2003spectra}. 
Under this model, $E[A]$ is $p_0*\textbf{1} \cdot\textbf{1}'$ for binary networks and $\lambda_0*\textbf{1} \cdot\textbf{1}'$ for count networks. 

%
The \textbf{R}ecursive \textbf{MAT}rix (R-MAT) model, introduced by \cite{chakrabarti2004r}, is different from other network generation models in one important aspect - we specify the number of edges, $M$, to assign to the network and then generate the network \citep{chakrabarti2004r,miller2015spectral}.
%
%
To assign these pre-specified number of edges, $M$, we start with an empty $n \times n$ adjacency matrix.
Next we set edge assignment probabilities $a, b, c, d$ such that 	$a > d > c = b$ and $a + b + c + d = 1$.
For illustration, suppose we have an empty $2 \times 2$ adjacency matrix and a single edge to assign, then the edge will be assigned to one of the four cells according to the the probability matrix
\[
\begin{bmatrix}
a       & b  \\
c       & d \\    
\end{bmatrix}.
\]
In general, for a given empty adjacency matrix $A$ of size $n \times n$, we subdivide the matrix into four partitions and randomly choose to assign the first edge to one of the partitions according to the above probability matrix.
Once the edge is designated to a partition, we subdivide that particular partition again into four sub-partitions and choose to assign that edge to one of the subdivisions based on the same probability matrix. This process is repeated iteratively until the edge ends up in a cell $A_{ij}$ of the adjacency matrix.
Since our network is undirected, the process is repeated for $M/2$ iterations to assign the $M$ edges to the adjacency matrix. Once cell $A_{ij}$ receives an edge, cell $A_{ji}$ also receives the edge. 
As we allow for self loops as in \cite{miller2015spectral}, an edge can end up in a cell $A_{ii}$.
Note that $n$ must be a power of two for the R-MAT model. 
In \cite{miller2015spectral}, the authors argued that since it is difficult to calculate $E[A]$ under the R-MAT model, it is better to use a ``rank-1 approximation'' such that 
\begin{equation} \label{eq_RMAT}
E[A] = \frac{\textbf{k} \textbf{k}^T}{2M},
\end{equation}
where $\textbf{k} = (k_1, \ldots, k_n)$ is the vector of observed degrees, and $M = \frac{1}{2}\sum k_i$.
We use the same approach in our paper for the sake of consistency.
Under the Chung Lu model \citep{chung2003spectra}, $p_{ij} = \frac{k_i k_j}{2M}$ where $k_1, \ldots, k_n$ are \textit{expected} degrees, i.e., popular nodes are more likely to interact with each other.
Following \cite{miller2015spectral}, in our paper the vector $(k_1, \ldots, k_n)$ in the Chung-Lu model for the binary case is set to be the observed degrees from the R-MAT randomly generated networks.
The use of the Chung-Lu model for the count network case is explained in section 6.
%
%
%
%
%


%
%
%

\section{Statistical evaluation of the chi-square algorithm}
In this section, we perform statistical evaluations of the chi-square algorithm by comparing the simulated test statistics to the chi-square distribution which is the implied distribution from \cite{miller2015spectral} and \cite{miller2010toward}. We also introduce some methodological improvements that will be useful to a practitioner implementing this algorithm.

\subsection{Chi-square algorithm methodology}

\begin{figure}[!hbt] 
  \centering 
  \includegraphics[width=75mm, keepaspectratio = true, scale=0.25]{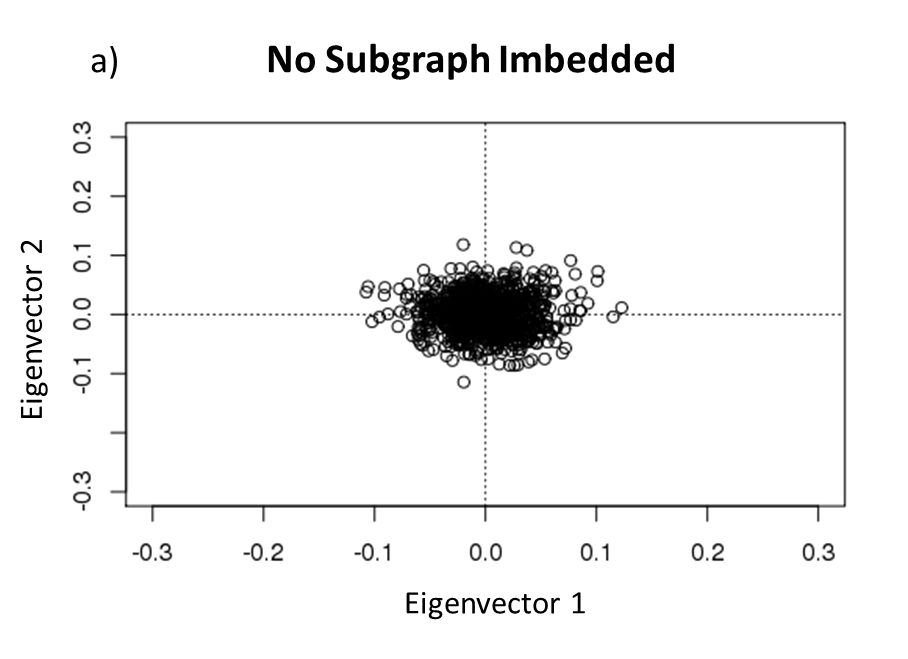} 
  \centering 
  \includegraphics[width=75mm, keepaspectratio = true, scale=0.25]{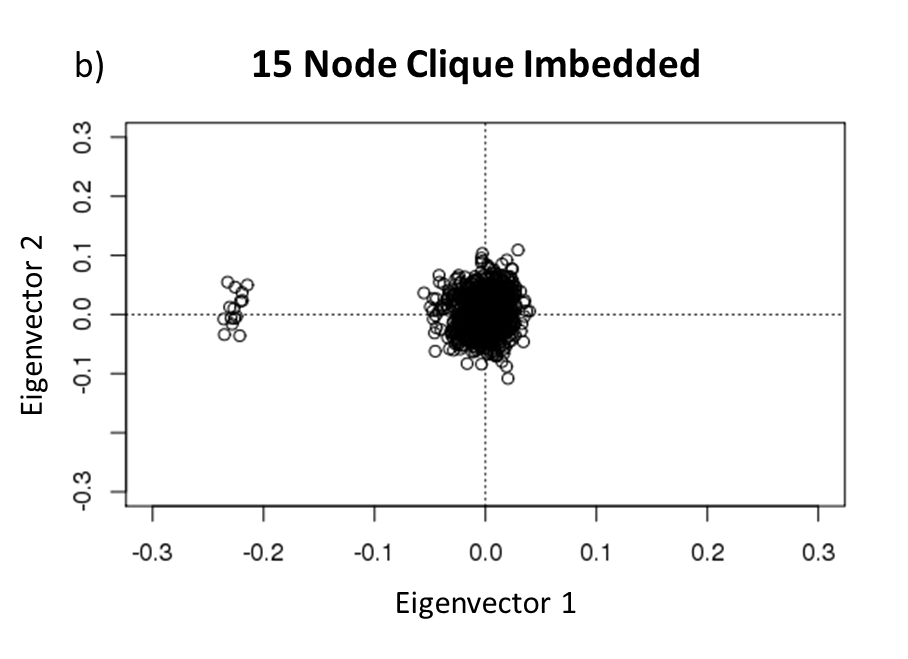} 
 \caption{ (a) ER Model with 1024 nodes with no anomalies showing radial symmetry about origin, (b) ER Model with 1024 nodes with anomalous sub-network present. Eigenvector one corresponds to the first principal component and eigenvector two corresponds to the second principal component. Background probability is $p_0$ = 0.1 and foreground probability is $p_1$ = 1 }
 \label{sub_nosub}
\end{figure}

For networks with no anomalies, empirical observations show that the first two principal components of the residual matrix, $B$, are radially symmetric.
%
The chi-square algorithm relies on this radial symmetry of the first two principal components to detect anomalies. 
One uses the number of points in each quadrant when plotting the first two principal components to calculate the detection statistic. 
For illustration, the left panel of Figure \ref{sub_nosub}  shows the first two principal components for a 1024-node ER network with no anomaly, whereas the right panel is with a 15 node clique embedded into the 1024-node ER network.
Note that the first two principal components are the eigenvectors corresponding to the two largest eigenvalues.
From Table 1 we observe that without an anomaly, the 1024 points are roughly uniformly distributed in the four quadrants, whereas in the anomalous case there is a substantially larger number of points in the fourth quadrant.

\begin{table}[h]
\begin{center}
\caption{Counts of points in each quadrant for Figure \ref{sub_nosub}}
 \label{sub_nosub_tab}
\begin{tabular}{ |c|c|c|c|c|c| } 
 \hline
 \textbf{Figure} & \textbf{Q1} & \textbf{Q2} & \textbf{Q3} & \textbf{Q4} & \textbf{Total}\\
 \hline
   (a) & 258 & 259 & 251 & 256 & \textbf{1024}\\ 
  \hline
  (b) & 250 & 247 & 254 & 273 & \textbf{1024} \\ \hline

\end{tabular}
\end{center}
\end{table}
For the chi-square algorithm, the first step involves obtaining the residual matrix of the network as described in Equation \eqref{eq_expected}. Then we obtain the two eigenvectors, $\textbf{X}_1$ and $\textbf{X}_2$ corresponding to the two largest eigenvalues of \textbf{B} and plot these orthogonal eigenvectors on a Cartesian coordinate system. Next we compute a $2 \times 2$ contingency table where each cell of the table is the number of points that fall in a particular quadrant. The $2 \times 2$ contingency table is a matrix $\textbf{O}$ with elements $O_{pq}$. We compute the expected number of points in each cell of the table assuming independence, yielding
\begin{equation} \label{eq_Chisqr}
\overline{O}_{pq}  = \frac{(O_{p1} + O_{p2})(O_{1q} + O_{2q})}{n}.
\end{equation}

The chi-square statistic is then 
\begin{equation} \label{eq_ChiSq_sum}
\chi^2([x_1 x_2])  = \sum_{p} \sum_{q} \frac{(O_{pq} - \overline{O}_{pq})^2}{\overline{O}_{pq}}  
\end{equation}

Because the non-anomalous  case is assumed to yield points that are radially symmetric, rotating the Cartesian plane should have no effect.
An anomaly could project the points in a certain direction, so the Cartesian plane is rotated to maximize the detection statistic, i.e.,
\begin{equation} \label{eq_ChiSq_Rot}
\chi^2_{max}  = \max_{\theta}  \chi^2 ([x_1 x_2] \begin{bmatrix}
cos \theta       & -sin \theta  \\
sin \theta       &  cos \theta \\    
\end{bmatrix}^T).
\end{equation}
An anomaly is signaled if $\chi^2_{max}$ is unusually large compared to the benchmark distribution.
In \cite{miller2015spectral} or \cite{miller2010toward}, the authors do not explicitly mention the parameters used in the benchmark distribution.

\begin{algorithm}[h]
\KwIn{Observed network}
\KwOut{Alert for any anomalous subgraph detected}

\nl Obtain the two eigenvectors that correspond to the two largest eigenvalues of the residual matrix;

\nl Plot the values of the two eigenvectors evaluated for each node and count the number of points in each quadrant;

 \nl  Standardize the observed counts and calculate the $\chi^2$ statistics using Equation \eqref{eq_Chisqr} ;
 
 \nl  Rotate the eigenvectors using Equation \eqref{eq_ChiSq_Rot} and calculate $\chi^2$ statistics;
 
  \nl  Store the maximum $\chi^2$ statistic value, $\chi^2_{max}$;
  
  \nl Signal if observed detection statistic crosses specified threshold, i.e, $\chi^2_{max} > K$;
 
    \caption{{\bf chi-square algorithm} \label{Chsq}}
\end{algorithm}

Two main concerns have to be addressed when implementing the chi-square statistic method in practice: 

\begin{itemize}
\item{What is the appropriate cut-off value $K$ such that the algorithm signals when $\chi^2_{max} > K$? }
\item{Counting the number of points in each quadrant has some limitations such as accounting for points that lie on an axis or at the origin.}

\end{itemize}

Counting of vertices in each quadrant to calculate a detection statistic has some limitations to be explored in section \ref{sec_imprv}. Furthermore, \cite{miller2015spectral} implied that the detection statistic follows the chi-square distribution for all network order and background probability combinations. This implies that the detection statistic is (a) independent of the network order and, (b) independent of the background probabilities. 
Also a signaling value, $K$, is not specified although this is a critical component for detecting an anomaly. A practitioner applying the algorithm would need to know at what threshold the method signals the presence of an anomaly. In the following sections we will investigate these concerns.

\subsection{Evaluating statistical properties of chi-square algorithm when there is no anomaly}
The success of an anomaly detection method depends on the following three criteria:
\begin{enumerate}
	\item When there is no anomaly, the statistic should follow the benchmark distribution.
	\item When there is no anomaly, false alarm rates should stay close to target values.
	\item In the presence of an anomaly, the algorithm should signal with a high probability.
\end{enumerate}

We focus on networks without an anomaly and carry out a systematic investigation of the first two criteria.
For criterion one, histograms and Q-Q plots are used as visual tools for comparing the empirical distributions of the detection statistics with the theoretical distributions. 
Further, note that a signal is raised when the statistic exceeds some pre-determined upper quantile of the theoretical distribution.
Therefore, for criterion two to hold, upper quantiles of the empirical distribution should closely match upper quantiles of the theoretical distribution.
To study this, we specifically compare the upper quantiles (95\% - 99\%) of the empirical distribution with those of the theoretical distribution. 

\cite{miller2015spectral} reported results only for networks with $n=4096$ nodes.
We found, however, that the performance of the anomaly detection methods can vary with the order of the network, the sparsity of the network, as well as the size and nature of the anomaly.
To reflect a wide variety of possible scenarios, we consider a broad range of networks, with $n = 128, 256, 512, 1024$ and also background probabilities, $p_0 = 0.01, 0.05, 0.1, 0.3$ for the ER model.
For the R-MAT model, we use $M=n(n-1)p_0$ for the $(n, p_0)$ combinations to generate networks with same order and density as the ER networks, but with edges assigned as per the R-MAT model. The base edge assignment matrix has the probability values; $a = 0.5, b = 0.125, c = 0.125, d = 0.25$.
For the Chung-Lu model, for each $(n, p_0)$ combination, we generate a single network from the R-MAT model and use the observed degrees $\mathbf{k} = (k_1, \ldots, k_n)$ as model parameters for the Chung-Lu model.
This approach for the three models is consistent with \cite{miller2015spectral}.
For brevity, we report a subset of the results in this section.
 Additional figures and tables are in the supplementary materials and show similar patterns.

\subsection{Comparing the test statistic to the chi-square distribution}
In \cite{miller2010toward}, the chi-square algorithm detection statistics is said to follow the gamma distribution with shape parameter two. The authors base this from empirical observations from 10,000 simulations for both the Erd\"os-R\'enyi and R-MAT model. The gamma distribution with shape parameter two is the chi-square distribution although the degrees of freedom, another parameter that is needed to generate the appropriate chi-square distribution, is not explicitly mentioned in either \cite{miller2010toward} or \cite{miller2015spectral}. It is implied in \cite{miller2010toward} that estimating this second parameter is dependent on the baseline model. Furthermore, these empirical observations were based on one network order, $n = 1024$. However, this makes the chi-square algorithm proposed in  \cite{miller2010toward} and \cite{miller2015spectral} impractical to a practitioner who intends to apply the algorithm to a static network. It is unclear if the detection statistic distribution depends on the network order, network model, or other attributes of the network. It is preferable, particularly when the intent is to apply the algorithm to a static network with no apriori information, that the detection algorithm has a baseline distribution that does not depend on the baseline network. To strengthen our claim, we use a $\chi^2$ with $df = 1$ in our paper as empirical observations show this is a good fit for the R-MAT model. We show in our statistical analysis to follow that the chi-square algorithm performance is inconsistent when applied to different network orders, network model, and network connectivities. 
%
%

\begin{figure*}[h]
	\begin{center}
		\includegraphics[height=90mm,width=1.0\linewidth, keepaspectratio=true]{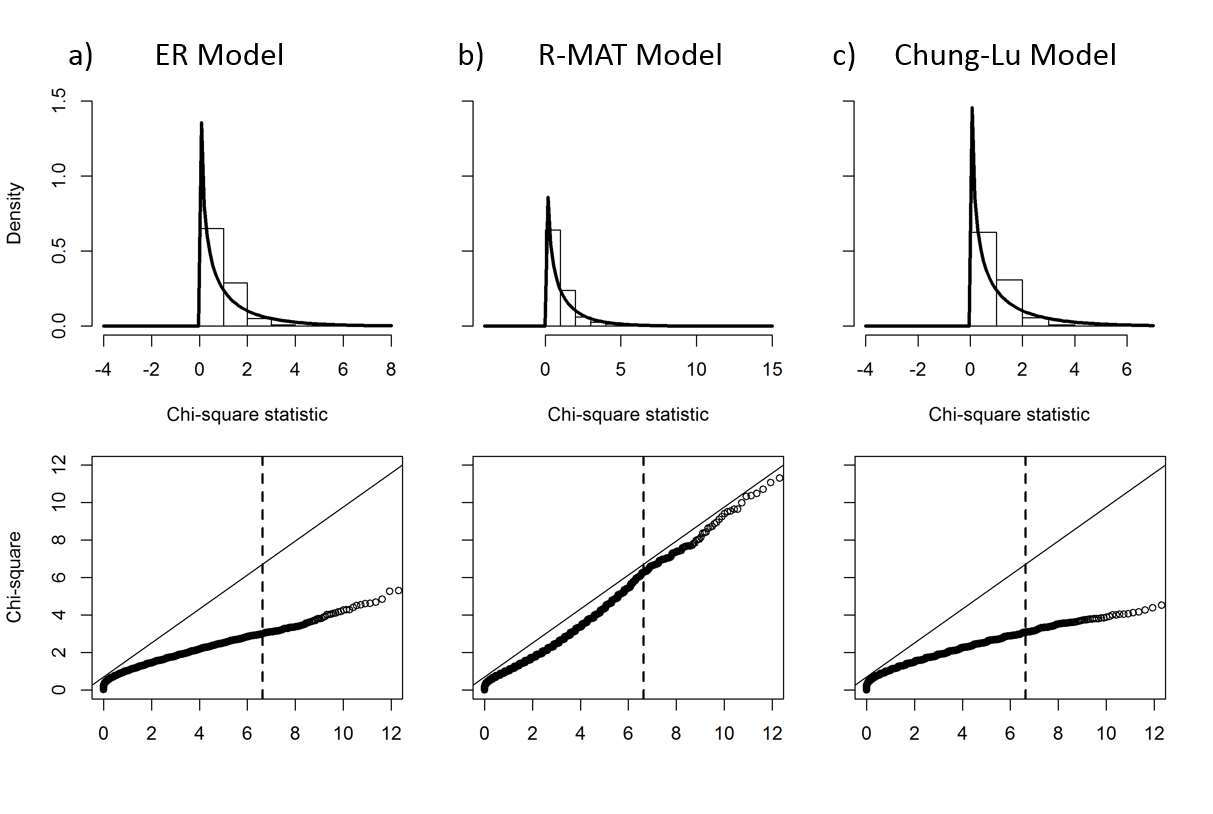}
		\caption{Top figures are histograms of the chi-square statistic based on 10,000 simulations with the chi-square distribution, $df = 1$, overlaid. $n = 512$ and $p_0 = 0.1$. Bottom figures are the corresponding Q-Q plots of the simulated chi-square statistics with the line $y = x$ representing the theoretical $\chi^2$ with $df=1$ and dashed line representing the $99^{th}$ percentile of the theoretical $\chi^2$ with $df=1$ distribution. ((a) Erd\"os-R\'enyi, (b) R-MAT, and (c) Chung-Lu Model) }
		\label{fig_Chi_all1}
	\end{center}
\end{figure*}

Selected results are shown in Figure \ref{fig_Chi_all1} and Table \ref{tab_10000Chisqr}; results from the other scenarios we explored are available in the supplementary material. 
%
For the ER and Chung-Lu models, we see that for all the network order combinations, histograms and Q-Q plots demonstrate that the algorithm statistic does not follow the chi-square distribution. 
%
In general, statistics based on the R-MAT model follow the chi-square distribution much better than for the ER and Chung-Lu models.

From Table \ref{tab_10000Chisqr}, it is further evident that the higher empirical quantiles do not match the theoretical $\chi^2$ quantiles in most cases.
The chi-square algorithm detection statistic is dependent on both the network order and background probability. 
In particular, sparse networks, $p_0 < 0.05$, have empirical quantiles much higher than the theoretical values. This again emphasizes our observation that the chi-square detection statistic does not follow the chi-square distribution and a detection value $K$ based on the chi-square theoretical distribution will yield unpredictable results. This is critical to a practitioner as setting the signaling threshold is dependent on the background connectivity of the observed network. Some recommendations for improving its performance is devising a better way to assign points to each quadrant, particularly for sparse networks. This improvement will be explored in section \ref{sec_imprv}.

\begin{table*}[h]
\small
\caption{Quantiles of the chi-square statistic. 10,000 non-anomalous simulations are run for the chi-square algorithm and the results compared to the quantiles of the chi-square statistic, $\chi^2$ with $df=1$.}
\label{tab_10000Chisqr}
\begin{adjustbox}{width=1\textwidth}
\centering
\begin{tabular}{|rr|rrrrr|rrrrr|rrrrr|}
\hline
\multicolumn{2}{|l|} {} &  \multicolumn{5}{|c|}{ER Model} & \multicolumn{5}{|c|}{R-MAT Model} &  \multicolumn{5}{|c|}{Chung-Lu Model} \\  \hline  
 Network order & $p_0$ & $95\%$ & $96\%$ & $97\%$ & $98\%$ & $99\%$ & $95\%$ & $96\%$ & $97\%$ & $98\%$ & $99\%$& $95\%$ & $96\%$ & $97\%$ & $98\%$ & $99\%$\\ 
  \hline

  128 & 0.010 & 57.92 & 59.69 & 61.25 & 63.87 & 67.66 & 36.11 & 37.27 & 38.55 & 40.11 & 43.18 & 20.97 & 21.70 & 22.50 & 23.58 & 25.39 \\ 
  128 & 0.050 & 3.72 & 3.94 & 4.28 & 4.73 & 5.49 & 4.97 & 5.28 & 5.65 & 6.23 & 7.14 & 4.98 & 5.19 & 5.52 & 5.92 & 6.53 \\ 
  128 & 0.300 & 2.18 & 2.31 & 2.44 & 2.77 & 3.16 & 1.99 & 2.11 & 2.34 & 2.57 & 3.16 & 2.05 & 2.30 & 2.38 & 2.68 & 3.02 \\ 
  256 & 0.010 & 22.98 & 24.17 & 25.80 & 27.86 & 31.63 & 37.04 & 38.08 & 39.26 & 40.76 & 43.73 & 25.02 & 25.79 & 26.68 & 28.20 & 30.68 \\ 
  256 & 0.100 & 2.60 & 2.72 & 2.94 & 3.24 & 3.74 & 2.70 & 2.90 & 3.18 & 3.51 & 4.10 & 2.79 & 2.91 & 3.15 & 3.47 & 3.93 \\ 
  256 & 0.300 & 2.17 & 2.31 & 2.45 & 2.68 & 3.19 & 2.36 & 2.60 & 2.89 & 3.22 & 3.86 & 2.17 & 2.33 & 2.53 & 2.74 & 3.22 \\ 
  512 & 0.010 & 9.76 & 10.44 & 11.39 & 12.76 & 15.05 & 30.03 & 30.97 & 32.09 & 33.89 & 36.66 & 21.13 & 21.86 & 23.16 & 24.59 & 26.97 \\ 
  512 & 0.100 & 2.40 & 2.56 & 2.76 & 3.08 & 3.44 & 3.08 & 3.33 & 3.66 & 4.12 & 5.29 & 2.67 & 2.82 & 3.02 & 3.33 & 3.81 \\ 
  512 & 0.300 & 2.17 & 2.31 & 2.48 & 2.69 & 3.07 & 3.33 & 3.60 & 4.06 & 4.89 & 6.35 & 2.27 & 2.34 & 2.52 & 2.75 & 3.15 \\ 
  1024 & 0.010 & 6.69 & 7.23 & 7.96 & 9.02 & 10.96 & 21.65 & 22.46 & 23.23 & 24.58 & 27.04 & 17.22 & 17.98 & 19.15 & 20.23 & 22.69 \\ 
  1024 & 0.100 & 2.28 & 2.42 & 2.58 & 2.81 & 3.18 & 3.97 & 4.44 & 5.20 & 6.25 & 8.52 & 2.60 & 2.76 & 3.01 & 3.30 & 3.83 \\ 
  1024 & 0.300 & 2.17 & 2.29 & 2.43 & 2.70 & 3.16 & 4.54 & 5.11 & 5.94 & 7.11 & 9.32 & 2.23 & 2.41 & 2.58 & 2.86 & 3.32 \\ 
   \hline
    \multicolumn{2}{|c|} {\textbf{$\chi^2$ with $df=1$}} & \textbf{3.84} & \textbf{4.22} & \textbf{4.71} & \textbf{5.41} & \textbf{6.63} & \textbf{3.84} & \textbf{4.22} & \textbf{4.71} & \textbf{5.41} & \textbf{6.63} & \textbf{3.84} & \textbf{4.22} & \textbf{4.71} & \textbf{5.41} & \textbf{6.63} \\ \hline
\end{tabular}
\end{adjustbox}
\end{table*}

\subsection{Improving the chi-square algorithm}\label{sec_imprv}
\begin{figure}[h] 
  \centering 
  \includegraphics[width=75mm, keepaspectratio = true, scale=0.25]{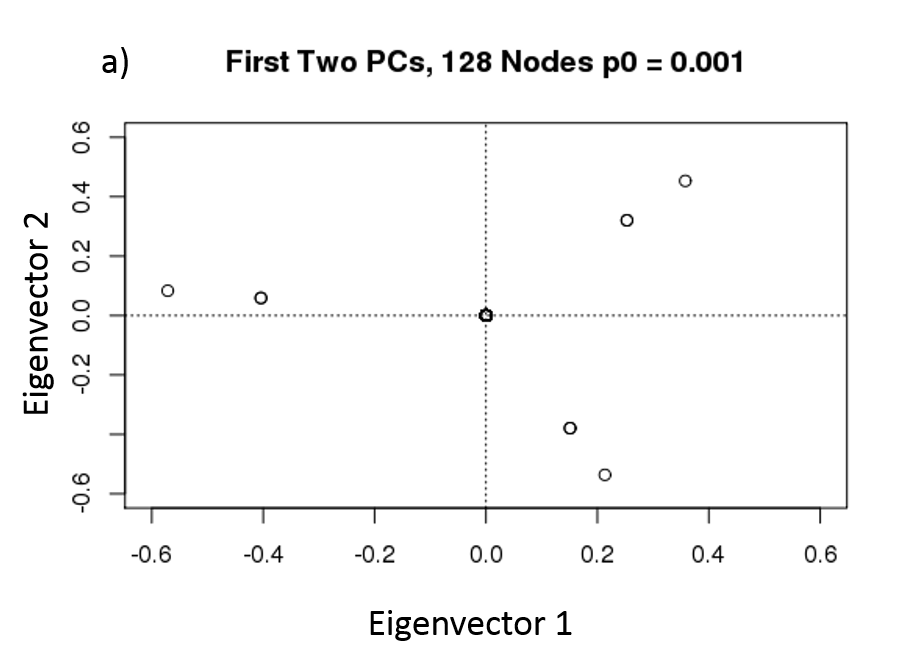} 
  \centering 
  \includegraphics[width=75mm, keepaspectratio = true, scale=0.25]{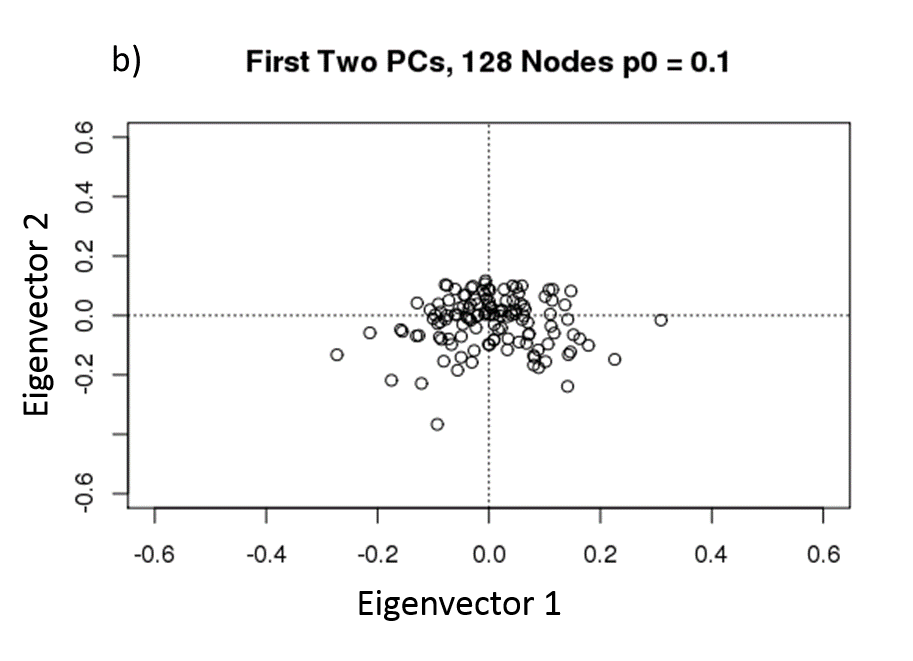} 
 \caption{ Figure (a) Sparse network with $n = 128$ and $p_0 = 0.001$, ER Model. There are 128 points in this plot of the first two principal components of the residual matrix although most are at the origin. Figure (b) Dense network with $p_0 = 0.1$ and we observe radial symmetry.}
 \label{PC128}
\end{figure}

\begin{figure}[h] 
  \centering 
  \includegraphics[width=75mm, keepaspectratio = true, scale=0.25]{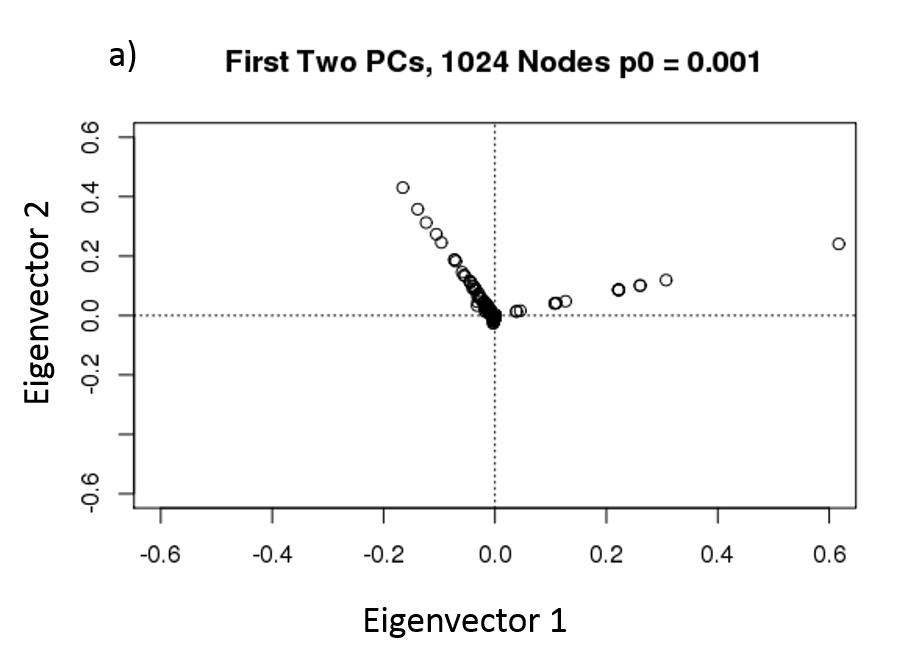} 
  \centering 
  \includegraphics[width=75mm, keepaspectratio = true, scale=0.25]{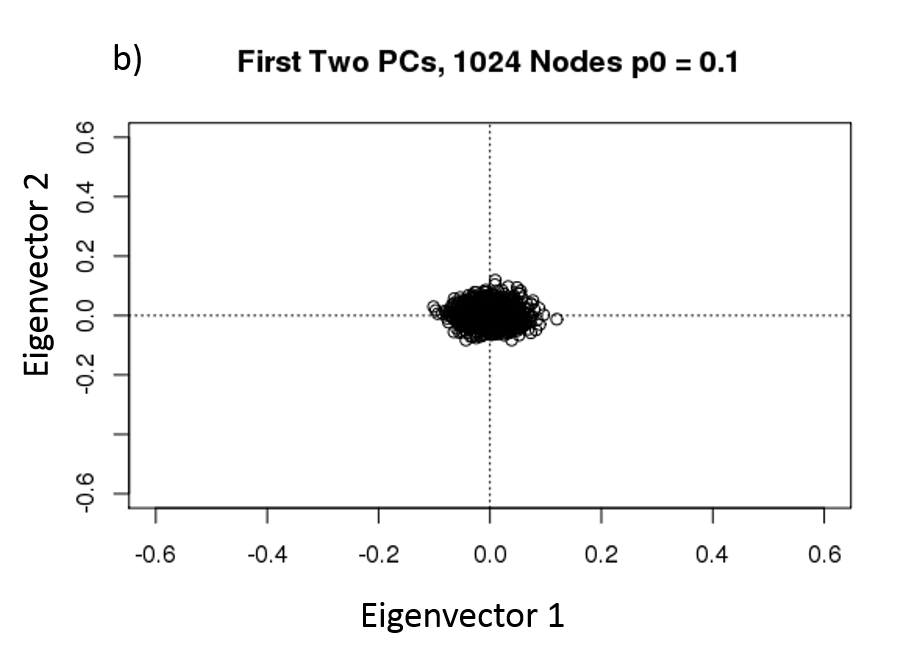} 
 \caption{ Figure (a) Sparse network with n = 1024 and $p_0 = 0.001$, ER Model. There are a total of 1024 points in this plot of the first two principal components of the residual matrix although most are centered at the origin.  Figure (b) Dense network with $p_0 = 0.1$ and we observe radial symmetry.}
 \label{PC1024}
\end{figure}

One of the noticeable concerns with the chi-square algorithm proposed in \cite{miller2015spectral} is its poor performance with sparse networks. We observed in Table \ref{tab_10000Chisqr} that the chi-square algorithm can yield very high statistic values for sparse networks. 
We note that sparse graphs are likely to be disconnected, which might be a possible reason for this phenomenon.
For $p_0 < 0.05$, the detection statistic quantiles are about an order of magnitude larger than the theoretical values. We hypothesize that this is due to how points are assigned to a quadrant. In sparse networks, the first two principal components of the residual matrix have a higher proportion of values close to zero. So when plotted, although radial symmetry is maintained, a significant number of points end up near or on the origin. Figures \ref{PC128} and \ref{PC1024} illustrate this phenomena. In these figures, some points are in fact on the origin but due to the computational limitations of some spectral decomposition calculations these values are actually approximations. One result of this is an abundance of points that end up in one particular quadrant. Hence, when assigning points to the $2 \times 2$ table as the algorithm requires, there is a tendency for a particular quadrant to be over-represented.  
As an example of how the quadrant count is affected, Table \ref{quad_128} and Table \ref{quad_1024} show the results when the graph is sparse versus when it is more connected. In Table \ref{quad_128} Q4 is over-represented and in Table \ref{quad_1024} Q3 is over-represented for the sparse network. Therefore a methodology is needed to account for points that end up on the origin or one of the axes. Note that there are no anomalies present in either of the examples below.

\begin{table}[h!]
\begin{center}
\caption{Counts of points in each quadrant}
\label{quad_128}
\begin{tabular}{ |c|c|c|c|c|c| } 
 \hline
 \textbf{$p_0$} & \textbf{Q1} & \textbf{Q2} & \textbf{Q3} & \textbf{Q4} & \textbf{Total}\\
 \hline
  0.001 & 5 & 27 & 6 & 90 & \textbf{128}\\ 
  \hline
  0.1 & 30 & 32 & 31 & 35 & \textbf{128} \\
  \hline
\end{tabular}
\end{center}
\end{table}

\begin{table}[h!]
\begin{center}
\caption{Counts of points in each quadrant}
\label{quad_1024}
\begin{tabular}{ |c|c|c|c|c|c| } 
 \hline
 \textbf{$p_0$} & \textbf{Q1} & \textbf{Q2} & \textbf{Q3} & \textbf{Q4} & \textbf{Total}\\
 \hline
  0.001 & 15 & 81 & 928 & 1& \textbf{1024}\\ 
  \hline
  0.1 & 246 & 238 & 285 & 255 & \textbf{1024} \\
  \hline
\end{tabular}
\end{center}
\end{table}

It should be noted that this behavior is network order dependent. That is, for the same background connectivity value, the plot of the first two principal components of a larger network tends to be relatively more compact as compared to a smaller network. We observe this in Figures \ref{PC128}  and \ref{PC1024}. To verify this, we ran multiple simulations with no anomalies present and observed that the distance of points from the origin is inversely proportional to the square root of the network order. In particular, $D $ $\propto$ $\frac{k}{\sqrt[]{n}}$. Also, we observed that this distance, $D$, is also inversely proportional to the square root of the connectivity of the graph, $p_0$, that is $D$ $\propto$ $\frac{k}{\sqrt[]{p_0}}$. This relationship is relatively weak, however, when compared to the effect network order has on the average distance of a point from the origin. 

This implies that we can improve the performance of the chi-square statistic by allocating points that are close to the origin equally to all four quadrants. We can do this by specifying that points that are within a certain distance $D_0$ from the origin should be approximately equally distributed to all four quadrants. This distance $D_0$ should be adjusted to compensate for the order of the network. In our improvement, we specify $D _0$ based on calculating the distances of every point from the origin. Using the relationship that $D = \frac{k}{\sqrt[]{n}}$, the best performing $k$ value that was observed through simulation results was $k$ = 0.35. This was the $k$ value that worked for the Erd\"os-R\'enyi, R-MAT, and Chung-Lu models. This approach also resolves one of the concerns with points lying  on an axis. Figures \ref{PC128}  and \ref{PC1024} and empirical observations showed that points a significant distance away from the origin rarely lie on one of the axes.

The top rows  of Table \ref{tab_10000Imprv} shows the simulation results for $\chi^2$ percentiles for both the Erd\"os-R\'enyi, R-MAT, and Chung-Lu models with no improvements made to the detection statistic. The bottom rows  of Table \ref{tab_10000Imprv} show the simulation results for both the Erd\"os-R\'enyi, R-MAT, and Chung-Lu models with our improved methodology. It is observed that for the improved version, the behavior of having significantly higher detection statistics than expected from the theoretical distribution is limited. This is most apparent for the R-MAT and Chung-Lu models. Overall, our modifications to the algorithm improve its performance substantially.

\begin{table*}[h]
\caption{Simulation results for the revised chi-square algorithm are compared to the theoretical chi-square distribution. Results only show the sparse networks based on $p_0 = 0.05$ when $n = 128$ and $p_0 = 0.01$ for other network orders. Includes both the statistics without any improvements, top rows, and algorithm results with improvement, bottom rows. Only the larger quantiles are presented in this table, i.e., $95^{th}$ to $99^{th}$ percentiles.  }
\label{tab_10000Imprv}
\begin{adjustbox}{width=1\textwidth}
\centering
\begin{tabular}{|rr|rrrrr|rrrrr|rrrrr|}
\hline
\multicolumn{2}{|l|} {\textbf{No improvements added }} &  \multicolumn{5}{|c|}{ER Model} & \multicolumn{5}{|c|}{R-MAT Model} &  \multicolumn{5}{|c|}{Chung-Lu Model} \\  \hline  
 Network order & $p_0$ & $95\%$ & $96\%$ & $97\%$ & $98\%$ & $99\%$ & $95\%$ & $96\%$ & $97\%$ & $98\%$ & $99\%$& $95\%$ & $96\%$ & $97\%$ & $98\%$ & $99\%$\\ 
  \hline

  128 & 0.050 & 3.72 & 3.94 & 4.28 & 4.73 & 5.49 & 4.97 & 5.28 & 5.65 & 6.23 & 7.14 & 4.98 & 5.19 & 5.52 & 5.92 & 6.53 \\ 
  256 & 0.010 & 22.98 & 24.17 & 25.80 & 27.86 & 31.63 & 37.04 & 38.08 & 39.26 & 40.76 & 43.73 & 25.02 & 25.79 & 26.68 & 28.20 & 30.68 \\ 
  512 & 0.010 & 9.76 & 10.44 & 11.39 & 12.76 & 15.05 & 30.03 & 30.97 & 32.09 & 33.89 & 36.66 & 21.13 & 21.86 & 23.16 & 24.59 & 26.97 \\ 
  1024 & 0.010 & 6.69 & 7.23 & 7.96 & 9.02 & 10.96 & 21.65 & 22.46 & 23.23 & 24.58 & 27.04 & 17.22 & 17.98 & 19.15 & 20.23 & 22.69 \\ \hline
   \multicolumn{2}{|c|} {\textbf{$\chi^2$ with $df=1$}} & \textbf{3.84} & \textbf{4.22} & \textbf{4.71} & \textbf{5.41} & \textbf{6.63} & \textbf{3.84} & \textbf{4.22} & \textbf{4.71} & \textbf{5.41} & \textbf{6.63} & \textbf{3.84} & \textbf{4.22} & \textbf{4.71} & \textbf{5.41} & \textbf{6.63} \\ \hline
   \hline \hline
\multicolumn{2}{|l|} {\textbf{Improvement added }} &  \multicolumn{5}{|c|}{ER Model} & \multicolumn{5}{|c|}{R-MAT Model} &  \multicolumn{5}{|c|}{Chung-Lu Model} \\  \hline  
   Network order & $p_0$ & $95\%$ & $96\%$ & $97\%$ & $98\%$ & $99\%$ & $95\%$ & $96\%$ & $97\%$ & $98\%$ & $99\%$& $95\%$ & $96\%$ & $97\%$ & $98\%$ & $99\%$\\ 
  \hline
  128 & 0.050 & 3.61 & 3.80 & 4.17 & 4.58 & 5.34 & 3.10 & 3.25 & 3.56 & 3.91 & 4.41 & 2.55 & 2.70 & 2.86 & 3.14 & 3.57 \\ 
  256 & 0.010 & 11.58 & 12.47 & 13.49 & 14.95 & 16.88 & 6.06 & 6.41 & 6.97 & 7.68 & 8.69 & 3.87 & 4.14 & 4.39 & 4.84 & 5.63 \\ 
  512 & 0.010 & 9.02 & 9.70 & 10.50 & 11.89 & 14.04 & 6.68 & 7.03 & 7.48 & 8.14 & 9.21 & 5.65 & 6.02 & 6.39 & 7.02 & 8.00 \\ 
  1024 & 0.010 & 6.43 & 7.06 & 7.75 & 8.77 & 10.45 & 6.59 & 6.91 & 7.34 & 8.03 & 9.06 & 8.47 & 9.11 & 9.75 & 10.65 & 12.28 \\ 
   \hline
    \multicolumn{2}{|c|} {\textbf{$\chi^2$ with $df=1$}} & \textbf{3.84} & \textbf{4.22} & \textbf{4.71} & \textbf{5.41} & \textbf{6.63} & \textbf{3.84} & \textbf{4.22} & \textbf{4.71} & \textbf{5.41} & \textbf{6.63} & \textbf{3.84} & \textbf{4.22} & \textbf{4.71} & \textbf{5.41} & \textbf{6.63} \\ \hline
\end{tabular}
\end{adjustbox}
\end{table*}

In Table 5, we used $k=0.35$ in the formula $D = \frac{k}{\sqrt[]{n}}$.
Since this value of $k$ was obtained from our simulations, it is important to investigate to what extent the improved method depends on the value of the tuning parameter, $k$.
To study this, we carried out a sensitivity analysis using $k = 01, 0.2, 0.3, 0.4, 0.5, 1$ and the same model settings as above.
The results are reported in Table 6.
We can see that for smaller values of $k$, e.g., $k = 0.1$, the ER model performs poorly, particularly for smaller networks. 
For all other values of $k$, the improvements gained from  assigning points equally to all quadrants are comparable to using $k=0.35$.
Therefore, in networks with no anomaly, the gain in performance from our proposed improvement is robust to values of $k$ in the range $(0.2, 1)$.

\begin{table*}[h]
\caption{Simulation results for the revised chi-square. Results only show the sparse networks based on $p_0 = 0.05$ when $n = 128$ and $p_0 = 0.01$ for other network sizes. Includes results for different values of the constant $k = 0.1, 0.2, 0.3, 0.4, 0.5$, and $1$.}
\label{tab_Imprv_K_0.1}
\begin{adjustbox}{width=1\textwidth}
\centering
\begin{tabular}{|rr|rrrrr|rrrrr|rrrrr|}
\hline
\multicolumn{2}{|l|} {\textbf{k = 0.1 }} &  \multicolumn{5}{|c|}{ER Model} & \multicolumn{5}{|c|}{R-MAT Model} &  \multicolumn{5}{|c|}{Chung-Lu Model} \\  \hline  
 Network size & $p_0$ & $95\%$ & $96\%$ & $97\%$ & $98\%$ & $99\%$ & $95\%$ & $96\%$ & $97\%$ & $98\%$ & $99\%$& $95\%$ & $96\%$ & $97\%$ & $98\%$ & $99\%$\\ 
  \hline
  128 & 0.050 & 60.06 & 60.06 & 60.06 & 60.06 & 60.06 & 2.71 & 2.80 & 3.05 & 3.16 & 3.37 & 2.77 & 2.98 & 3.19 & 3.36 & 3.94 \\ 
  256 & 0.010 & 15.30 & 16.89 & 20.33 & 124.03 & 124.03 & 4.27 & 4.43 & 4.71 & 5.25 & 6.52 & 3.52 & 3.74 & 3.95 & 4.17 & 4.71 \\ 
  512 & 0.010 & 8.63 & 9.16 & 10.11 & 11.09 & 13.32 & 4.42 & 4.78 & 5.10 & 5.77 & 6.51 & 2.67 & 2.78 & 2.98 & 3.18 & 3.63 \\ 
  1024 & 0.010 & 6.65 & 7.06 & 7.51 & 9.37 & 11.29 & 7.53 & 8.09 & 8.95 & 9.91 & 10.97 & 8.91 & 9.30 & 10.39 & 11.22 & 12.40 \\ 
     \hline
     \multicolumn{2}{|l|} {\textbf{k = 0.2 }} &  \multicolumn{5}{|c|}{ER Model} & \multicolumn{5}{|c|}{R-MAT Model} &  \multicolumn{5}{|c|}{Chung-Lu Model} \\  \hline  
  128 & 0.050 & 3.84 & 4.09 & 4.85 & 5.68 & 8.70 & 2.82 & 2.95 & 3.19 & 3.51 & 3.88 & 2.66 & 2.74 & 2.86 & 3.16 & 3.64 \\ 
  256 & 0.010 & 12.41 & 13.05 & 14.14 & 15.47 & 18.64 & 4.65 & 4.92 & 5.06 & 5.44 & 6.14 & 4.35 & 4.51 & 4.77 & 5.13 & 5.86 \\ 
  512 & 0.010 & 9.32 & 9.92 & 10.66 & 11.93 & 13.65 & 6.22 & 6.41 & 6.98 & 7.46 & 8.72 & 4.91 & 5.44 & 5.93 & 6.66 & 7.28\\ 
  1024 & 0.010 & 6.22 & 6.81 & 7.84 & 8.78 & 10.52 & 6.35 & 6.90 & 7.51 & 8.46 & 9.78 & 6.77 & 7.19 & 7.68 & 9.06 & 11.77 \\ 
  \hline
       \hline
     \multicolumn{2}{|l|} {\textbf{k = 0.3 }} &  \multicolumn{5}{|c|}{ER Model} & \multicolumn{5}{|c|}{R-MAT Model} &  \multicolumn{5}{|c|}{Chung-Lu Model} \\  \hline
128 & 0.050 & 3.69 & 3.95 & 4.23 & 4.58 & 5.00 & 2.95 & 3.11 & 3.49 & 3.67 & 4.27 & 2.56 & 2.74 & 2.88 & 3.36 & 3.92\\ 
   256 & 0.010 & 12.55 & 13.36 & 14.28 & 15.71 & 19.43 & 4.11 & 4.34 & 4.68 & 5.08 & 5.43 &  4.29 & 4.61 & 4.90 & 5.24 & 5.89  \\ 
  512 & 0.010 & 9.25 & 10.43 & 11.91 & 12.66 & 15.01 & 4.28 & 4.40 & 4.63 & 5.03 & 6.73 & 5.26 & 5.73 & 6.25 & 7.03 & 8.15 \\ 
  1024 & 0.010 & 6.05 & 6.69 & 7.27 & 8.26 & 10.17 & 6.50 & 6.98 & 7.40 & 8.71 & 10.52 & 8.00 & 8.73 & 9.12 & 9.89 & 12.10\\ 
   \hline
         \hline
     \multicolumn{2}{|l|} {\textbf{k = 0.4 }} &  \multicolumn{5}{|c|}{ER Model} & \multicolumn{5}{|c|}{R-MAT Model} &  \multicolumn{5}{|c|}{Chung-Lu Model} \\  \hline
128 & 0.050 & 3.58 & 3.90 & 4.09 & 4.52 & 4.93 & 2.41 & 2.70 & 3.02 & 3.13 & 3.61 & 2.74 & 2.91 & 3.15 & 3.56 & 4.17\\ 
256 & 0.010 & 12.61 & 13.36 & 13.96 & 15.29 & 19.29 & 4.18 & 4.51 & 4.75 & 5.50 & 6.14 & 3.78 & 3.87 & 4.26 & 4.56 & 5.11\\ 
  512 & 0.010 & 9.90 & 10.96 & 11.33 & 12.60 & 14.34 & 4.93 & 5.13 & 5.47 & 6.00 & 6.94 & 4.56 & 4.97 & 5.34 & 5.97 & 6.66 \\ 
  1024 & 0.010 & 6.13 & 6.49 & 7.22 & 7.91 & 9.73 & 8.00 & 8.53 & 8.94 & 9.52 & 10.82 & 7.00 & 7.27 & 7.76 & 8.59 & 10.83 \\ 
           \hline
     \multicolumn{2}{|l|} {\textbf{k = 0.5 }} &  \multicolumn{5}{|c|}{ER Model} & \multicolumn{5}{|c|}{R-MAT Model} &  \multicolumn{5}{|c|}{Chung-Lu Model} \\  \hline
    128 & 0.050 & 3.62 & 3.79 & 4.02 & 4.34 & 4.99 & 2.67 & 2.75 & 3.09 & 3.34 & 3.83 & 2.67 & 2.95 & 3.24 & 3.57 & 4.06 \\ 
  256 & 0.010 & 12.09 & 12.63 & 13.56 & 16.01 & 17.66 & 4.04 & 4.52 & 4.88 & 5.19 & 6.09 & 3.91 & 4.15 & 4.35 & 4.76 & 5.98 \\ 
  512 & 0.010 & 8.63 & 9.16 & 10.39 & 11.85 & 15.02 & 5.10 & 5.36 & 5.80 & 6.51 & 7.58 & 4.87 & 5.05 & 5.45 & 6.18 & 7.10 \\ 
  1024 & 0.010 & 6.49 & 7.20 & 8.20 & 9.18 & 11.30 & 8.17 & 8.43 & 9.33 & 10.13 & 11.51 & 6.21 & 6.60 & 7.10 & 7.95 & 9.07 \\ \hline
      \multicolumn{2}{|l|} {\textbf{k = 1 }} &  \multicolumn{5}{|c|}{ER Model} & \multicolumn{5}{|c|}{R-MAT Model} &  \multicolumn{5}{|c|}{Chung-Lu Model} \\  \hline
 128 & 0.050 & 3.39 & 3.73 & 4.03 & 4.61 & 5.22 & 2.41 & 2.66 & 2.78 & 3.14 & 4.04 & 2.66 & 2.72 & 2.85 & 3.21 & 3.62\\ 
  256 & 0.010 & 11.45 & 12.09 & 12.71 & 15.06 & 16.31 & 4.77 & 5.08 & 5.51 & 5.92 & 7.33 & 3.82 & 4.08 & 4.29 & 4.80 & 5.47\\ 
  512 & 0.010 & 9.41 & 10.07 & 10.28 & 11.19 & 13.19 & 5.00 & 5.25 & 5.68 & 6.20 & 7.00 & 5.98 & 6.27 & 6.59 & 7.06 & 8.03\\ 
  1024 & 0.010 & 6.65 & 7.08 & 7.91 & 8.55 & 10.15 & 8.75 & 9.30 & 9.81 & 10.47 & 12.22 & 6.69 & 7.33 & 7.77 & 8.32 & 9.18 \\ \hline
    \multicolumn{2}{|c|} {\textbf{$\chi^2$ with $df=1$}} & \textbf{3.84} & \textbf{4.22} & \textbf{4.71} & \textbf{5.41} & \textbf{6.63} & \textbf{3.84} & \textbf{4.22} & \textbf{4.71} & \textbf{5.41} & \textbf{6.63} & \textbf{3.84} & \textbf{4.22} & \textbf{4.71} & \textbf{5.41} & \textbf{6.63} \\ \hline
 \hline
\end{tabular}
\end{adjustbox}
\end{table*}

\subsection{Comparing performance of chi-square algorithm for both cases, no improvement and improvement added}
We investigated the performance of both cases with respect to false alarm rates, i.e., falsely signaling an anomaly when there is no anomaly, and detection rates, i.e., correctly signaling an anomaly when there is an anomaly.
We consider two evaluation metrics in this section.

\begin{enumerate}
		\setlength{\itemsep}{-0.4ex}
		\item False alarm rate (FAR) is $P[\text{signal }| \text{ no anomaly}]$, i.e., the proportion of cases where no anomaly is present, but the detection rule incorrectly signals an anomaly.

	\item Detection rate (DR) is $P[\text{signal } |\text{ anomaly}]$, i.e., the proportion of cases where anomaly is present, and the detection rule correctly signals an anomaly.
\end{enumerate} 
We have
\[
DR = \frac{TP}{TP + FN}, \; \; FAR = \frac{FP}{FP + TN},
\]
where the confusion matrix for calculating the detection rate (DR) and false alarm rate (FAR) is shown in Table \ref{det_False}.
As before, we carried out a systematic study using a broad range of simulation settings with the three network models.
We considered network orders $n =$ 128, 256, 512, 1024. For $n = 128$, the background connectivity was chosen to be $p_0$ = $0.05$ for other network orders, $p_0$ = $0.01$. Selecting a higher background connectivity of $p_0$ = $0.05$ for the network of order $n = $ 128 ensures that the majority of the nodes are indeed connected. If $p_0$ = $0.01$, then the average degree of the network would be $1.28$ which would result in a network with isolated nodes. For $n = 128$ and $n= 256$, we randomly embedded cliques of size 3\%, 4\%, 5\%, and 6\% of the network order. For $n = $ 512 and 1024 network orders, we randomly embedded cliques of size 1\%, 2\%, 3\%, and 4\% of the network order. For brevity, only the results for $n = 512$ are shown in this section as the other network orders led to similar results. Each detection and false alarm rate estimation was performed for the case where $\alpha = 0.05$, i.e., target FAR is $5\%$. 
We ran 500 simulations for each parameter combination of which 250 were with the anomaly and 250 were without the anomaly.

\begin{table}[ht]
\centering
\caption{Confusion Matrix}
\label{det_False}
\begin{tabular}{|l|l|l|}
\hline
 & Anomaly present  & No anomaly present \\ 
  \hline
Anomaly signaled & True Positive (TP) & False Positive (FP)  \\ \hline
Anomaly not signaled  & False Negative (FN) & True Negative (TN) \\ \hline
   \hline
\end{tabular}
\end{table}

We compared the performance of the chi-square algorithm with the revised algorithm. Table \ref{tab_Rev_Normal} shows that the improved chi-square algorithm retains the same detection power while significantly reducing the false alarm rates. This is most apparent with the R-MAT and Chung-Lu models. The false alarm rates are still much too high, however, so we recommend our modification of the $L_1$ norm algorithm instead in section \ref{L1sec}.

   \begin{table*}[h]
\caption{Detection and False Alarm Rates. Background probability, $p_0 = 0.01$ and $n = 512$. Foreground probability is $p_1 = 1$. We performed 500 simulations for each row with an anomalous subgraph randomly embedded in 250 of 500 simulations}
\label{tab_Rev_Normal}
\begin{adjustbox}{width=1\textwidth}
\small\begin{tabular}{|r|l|l|l|l|l|}
  \hline
 \multicolumn{2}{l} {ER Model} &  \multicolumn{2}{c}{Detection Rate \%} & \multicolumn{2}{c}{False Alarm Rate \%}\\  \hline  
 Network order & Subgraph Size & $\chi^2$ Ther. 95\%  & $\chi^2$ Revised 95\% & $\chi^2$ Ther. 95  & $\chi^2$ Revised 95  \\
  \hline
512 & 5 & 100.00 & 100.00 & 60.80 & 57.20 \\ 
  512 & 10 & 100.00 & 100.00 & 60.80 & 57.20 \\ 
  512 & 15 & 100.00 & 100.00 & 60.80 & 57.20 \\ 
  512 & 20 & 100.00 & 100.00 & 60.80 & 57.20 \\ 
   \hline
 \multicolumn{2}{l} {R-MAT Model} &  \multicolumn{2}{|c|}{Detection Rate} & \multicolumn{2}{|c|}{False Alarm Rate}\\  \hline  
  Network order & Subgraph Size & $\chi^2$ Ther. 95\%  & $\chi^2$ Revised 95\%  & $\chi^2$ Ther. 95\% & $\chi^2$ Revised 95\% \\
  \hline
  512 & 5 & 99.60 & 29.20 & 99.30 & 23.10 \\ 
  512 & 10 & 100.00 & 100.00 & 99.30 &  23.10 \\ 
  512 & 15 & 100.00 & 100.00 & 99.30 &  23.10 \\ 
  512 & 20 & 100.00 & 100.00 & 99.30 &  23.10 \\ 
   \hline
   \multicolumn{2}{l} {Chung-Lu Model} &  \multicolumn{2}{|c|}{Detection Rate} & \multicolumn{2}{|c|}{False Alarm Rate}\\ \hline  
  Network order & Subgraph Size & $\chi^2$ Ther. 95\%  & $\chi^2$ Revised 95\%  & $\chi^2$ Ther. 95\% & $\chi^2$ Revised 95\%  \\ \hline
  512 & 5 & 100.00 & 58.80 & 99.80 & 9.70 \\ 
  512 & 10 & 100.00 & 100.00 & 99.80 &  9.70 \\ 
  512 & 15 & 100.00 & 100.00 & 99.80 &  9.70 \\ 
  512 & 20 & 100.00 & 100.00 & 99.80 &  9.70 \\ 
  \hline
\end{tabular}
\end{adjustbox}
\end{table*}

\section{Eigenvector L\textsubscript{1} norm algorithm methodology} \label{L1sec}
The $L_1$ norm of a vector \textbf{X}  = ($x_1$, $x_2$,..., $x_n$) is defined as 
$|\textbf{X}|_1 = \sum_{i=1}^{n} |x_i| $.
In \cite{miller2010L1norm} and \citep{miller2015spectral}, the authors proposed an anomaly detection technique based on $L_1$ norms of eigenvectors of the residual matrix $\mathbf{B}$.
Let $\mathbf{X}_1, \ldots, \mathbf{X}_m$ be the orthonormal eigenvectors corresponding to the $m$ largest eigenvalues of $\mathbf{B}$.
Heuristically,  when an anomaly is present, for some $k \in \{1, \ldots, m\}$, the elements of the vector $\textbf{X}_k$ that correspond to anomalous nodes will have absolute values that are significantly larger than other elements of the vector $\textbf{X}_k$.
Let $\textbf{X}_k = (x_1, \ldots, x_n)$.
Since the eigenvector is orthonormal,  $\sum_{i=1}^{n} x_i^2 = 1$ needs to be satisfied. 
In the presence of an anomaly, only a small portion of the elements, say $\{x_1, \ldots, x_s\}$ (where $s$ is much smaller than $n$), have larger values such that $\sum_{i=1}^{s} x_i^2 \approx 1$, whereas in the absence of an anomaly, $\{x_1, \ldots, x_n\}$ have approximately equal values (up to random variation).
 Therefore, the $L_1$ norm of $\textbf{X}_k$ will be significantly smaller when there is an anomaly, compared to when there is no anomaly.
 Consequently, an anomaly can be detected by low values of the $L_1$ norms of eigenvectors of $\mathbf{B}$.
 Formally, the $L_1$ norm  statistic $L$ is calculated as  
\begin{equation} \label{eq_L1min}
L = -\min_{1 \leq k \leq m}  \frac{|\textbf{X}_k|_1 - \mu_k}{\sigma_k}
\end{equation}
where $|\pmb{X}_k|_1$ is the $L_1$ norm of the $k^{th}$eigenvector, $\mu_k$ and $\sigma_k$ are the mean and standard deviation of $|\pmb{X}_k|_1$ when there is no anomaly, and $m$ is the number of eigenvalues used for anomaly detection.
Large values of $L$ indicate the presence of an anomaly.

Three issues present themselves immediately from the formulation.
First, the statistic needs to be standardized using the parameters $\mu_k, k = 1, \ldots, m$ and $\sigma_k, k = 1, \ldots, m$.
\cite{miller2015spectral} proposed to estimate these parameters from historical networks where no anomaly is present.
For each historical network observation of size $n$, where no anomaly is present, its residual matrix as in Equation \eqref{eq_expected}  is first calculated. Then for each residual matrix, an arbitrary set number of $m$ largest eigenvalues, where $m$  $\leq$ $n$, are sorted in decreasing order and the $L_1$ norms of the corresponding eigenvectors calculated. That is, an $L_1$ norm value is calculated for each eigenvector $\textbf{X}_k$ where $k$ = 1,2,...,$m$ and the corresponding eigenvalues, $\xi_1$ $\geq$ $\xi_2$,... $\geq$ $\xi_m$. Then the mean of the historically observed $L_1$ norms for each of the eigenvectors $\textbf{X}_k$'s is estimated, yielding $\hat\mu_k$ where $k$ = 1,2,...,$m$ along with their standard deviations $\hat\sigma_k$. When a new graph is observed, its $m$ largest eigenvalues are extracted in decreasing order and their corresponding eigenvector $L_1$ norms calculated. 

However, for a static network for which this method is intended, there are no historical networks available, as the only data are from a single snapshot of a network and it is not known whether there is an anomaly in that data.
Therefore a critical shortcoming of this method as formulated in \cite{miller2015spectral} is that it cannot be used for static networks.
In section 5 we propose a solution to this issue by using results from Extreme Value Theorem.

Second, the choice of $m$ is not clear, one could use all eigenvectors, i.e., $m=n$, or a smaller number $m < n$.
\citep{miller2015spectral} makes some recommendations, but the implications are not rigorously tested.
In sections 4 and 5, we carry out analyses involving the various choices of $m$ to resolve this issue.

Third, in practice one needs a benchmark distribution to decide whether the value of the statistic is large enough to signal an anomaly.
The authors suggest using the Gumbel distribution, but it is not fully clear why this is the benchmark distribution.
Further, even if one accepts the use of the Gumbel distribution, the authors do not specify what parameters should be used for the Gumbel distribution.
We propose two options for estimating the parameters, as outlined below.

The Gumbel distribution is defined by two parameters, the location parameter $a_m$ and the scaling parameter $b_m$ \citep{nadarajah2004beta}. Given that the eigenvectors of the residual matrix  follow a standard normal distribution, as we assume in our case, the parameters $a_m$ and $b_m$ can be calculated from the Extreme Value Theorem using
\begin{equation} \label{eq_am}
a_m  = -\Phi^{-1}(1/m), \;  b_m  = \frac{1}{a_m}, 
\end{equation}
%
%
%
where $\Phi$ is the cumulative density function of the standard normal distribution and $m$ is the number of random variables from which the extrema is derived. In our case, $m$ is the number of eigenvectors used for anomaly detection.

The parameters $a_m$ and $b_m$ can also be estimated using the Method of Moments (MOM) estimators which requires using historical data. In this case,
\begin{equation} \label{eq_am_MOM}
\hat{a}_m  = \frac{1}{h} \sum_{i=1}^{h} L_{i} - \hat{b}_m\gamma, \hat{b}_m  = \frac{\sqrt[]{6}S}{\pi}, 
\end{equation}
%
%
where $h$ is the number of historical networks, $L_i$ is the $L_1$ norm detection statistic for network(i), $\gamma$ $\approx$ 0.57722, $S$ is the standard deviation of the $L_1$ norm detection statistics from the $h$ historical observations.
In subsequent sections we study the performance of the method using both the MOM estimators and the Extreme Value Theorem estimators.

\begin{algorithm}[h]
\KwIn{Observed network}
\KwOut{Alert for any anomalous subgraph detected}

\nl Obtain $\hat\mu_k$ and $\hat\sigma_k$ from historical networks;

\nl Standardize observed network $L_1$ norms for each eigenvector with corresponding $\hat\mu_k$ and $\hat\sigma_k$;

 \nl  Calculate location parameter $a_m$ and scaling parameter $b_m$;
 
 \nl  Transform standardized observed $L_1$ norms to standard Gumbel distribution using parameters $a_m$ and $b_m$;
 
  \nl Signal if observed network detection statistic for a given eigenvector crosses a specified threshold, $L_i > K$;
    \caption{{\bf $L_1$ norm algorithm} \label{L1norm}}
\end{algorithm}

%

The three issues we have identified can significantly impact a practitioner's ability to implement the algorithm as will be demonstrated in sections 4 and 5.  These are also limitations not stated explicitly in \cite{miller2015spectral} as the authors claimed the algorithm is applicable to a static observed network with no apriori information. Also, a criteria for signaling is not explicitly presented in \cite{miller2015spectral}. In following sections we further elaborate on possible signal detection thresholds and the resulting relative performance. 

\subsection{Statistical Properties of Eigenvector L\textsubscript{1} norm algorithm}
In \cite{miller2015spectral}, the detection statistic from the $L_1$ norm algorithm is stated to follow a Gumbel distribution. This distribution depends on two parameters, the location  and scaling parameters $a_m$ and $b_m$, respectively. These parameters need to be estimated in order to standardize the observed detection statistic. Furthermore, the effect of the number of eigenvectors, $m$, on the detection statistic result is not discussed in \cite{miller2015spectral}. An arbitrary value, $m$ = 100, is used without a discussion or validation of the approach. In this section, we will compare two different estimation techniques for $a_m$ and $b_m$ where in one case we use the Method of Moments estimator (MOM) based on historical data to estimate these parameters as in Equation \eqref{eq_am_MOM} and in the second case we use the Extreme Value Theorem approach in Equation \eqref {eq_am}. We also studied the effect of the arbitrarily set value of $m$ on the non-anomalous behavior of the $L_1$ norm statistic, by setting $m < n$ in one case and $m$ = $n$ in another. If the algorithm statistic follows the Gumbel distribution, then we should expect better performance when $m = n$ as the $(a_m, b_m)$ estimates should be more accurate with a larger sample size. We also used the same range of simulation settings and evaluation metrics as we did for the chi-square algorithm in section 3. 

\subsubsection {Estimating $a_m$ and $b_m$ using historical data and setting $m < n$}
\begin{figure*}[h]
	\begin{center}
		\includegraphics[height=90mm,width=1.0\linewidth, keepaspectratio=true]{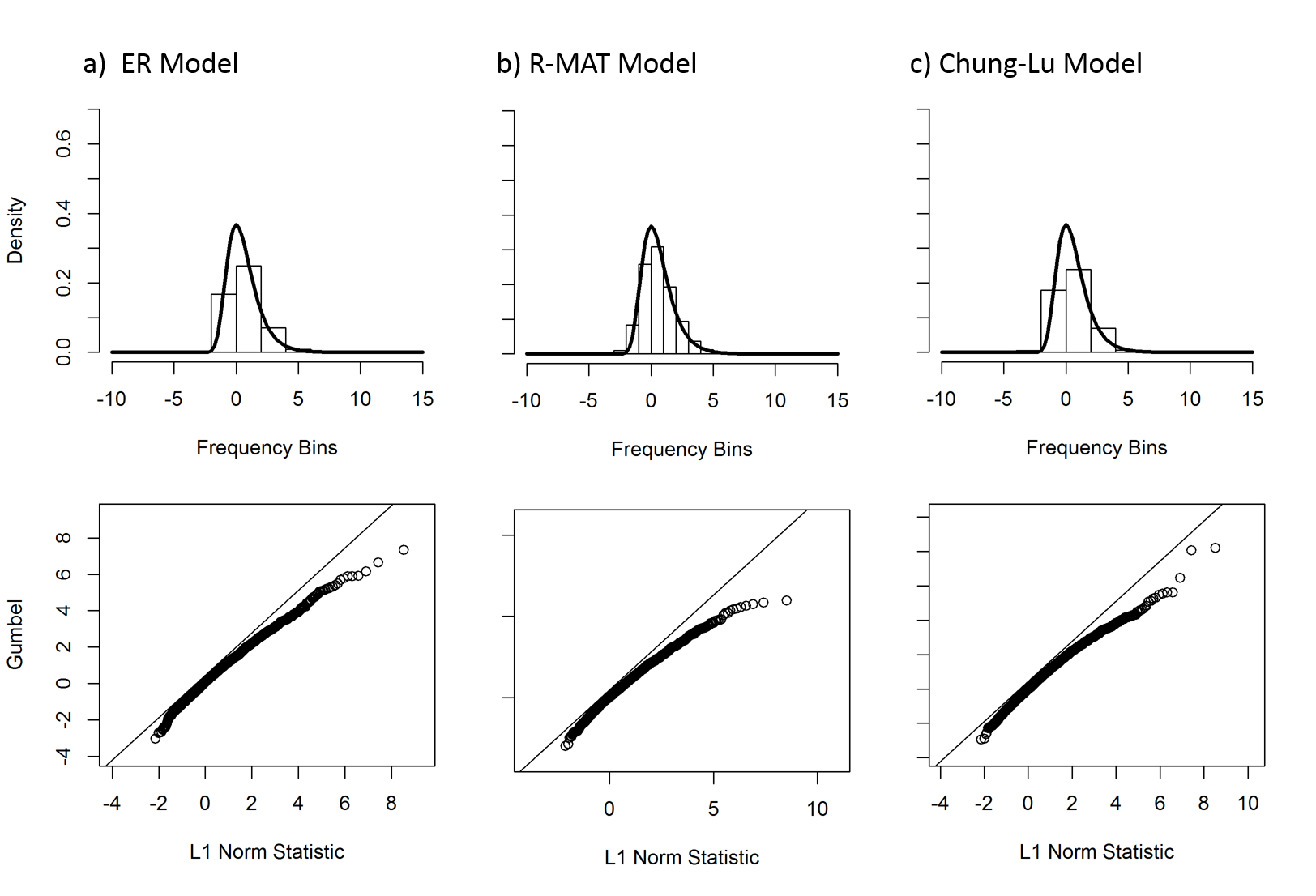}
		\caption{Top figures are histogram density plots when parameters $a_m$ and $b_m$ are estimated using historical data and MOM estimators with $m < n$. Solid black line represents the theoretical Gumbel distribution. Bottom figures are the corresponding Q-Q plots of the simulated $L_1$ norm statistics with the line $y = x$ representing the theoretical Gumbel distribution and dashed line representing the $99^{th}$ percentile of the theoretical Gumbel distribution. This example is with $n = 512$ and $p_0 = 0.1$; other scenarios are presented in the supplementary material and follow similar patterns. ((a) Erd\"os-R\'enyi, (b) R-MAT, and (c) Chung-Lu Model) }
		\label{fig_HistER_L1m512}
	\end{center}
\end{figure*}

First, we consider the $m<n$ case by setting $m=30$ for $n=128$ and $n=256$, and $m=50$ for $n=512$ and $n= 1024$. We use the MOM estimator as in Equation \eqref{eq_am} to estimate $a_m$ and $b_m$ using historical data. This involves first generating 1000 random networks as historical data 
to estimate $\mu_k$, $\sigma_k$ for $k=1, \ldots, m$. We used a very large number of historical networks in order to obtain a bound on the performance of the method in practice.
The results are shown in Figure \ref{fig_HistER_L1m512}.
For the Erd\"os-R\'enyi model, the empirical distribution is similar to the theoretical Gumbel distribution, although they are dissimilar at the higher quantiles. The same observation is noted in the plot comparisons for the R-MAT and Chung-Lu model. Deviations at the higher quantiles reduces the usefulness of the algorithm to a practitioner because it makes setting an effective signaling threshold more difficult. 
Table \ref{tab_10000AllL1m} also corroborates our conclusions for this case. 


\subsubsection{Estimating $a_m$ and $b_m$ using the Extreme Value Theorem and setting $m < n$}
\begin{figure*}[t]
	\begin{center}
		\includegraphics[height=90mm,width=1.0\linewidth, keepaspectratio=true]{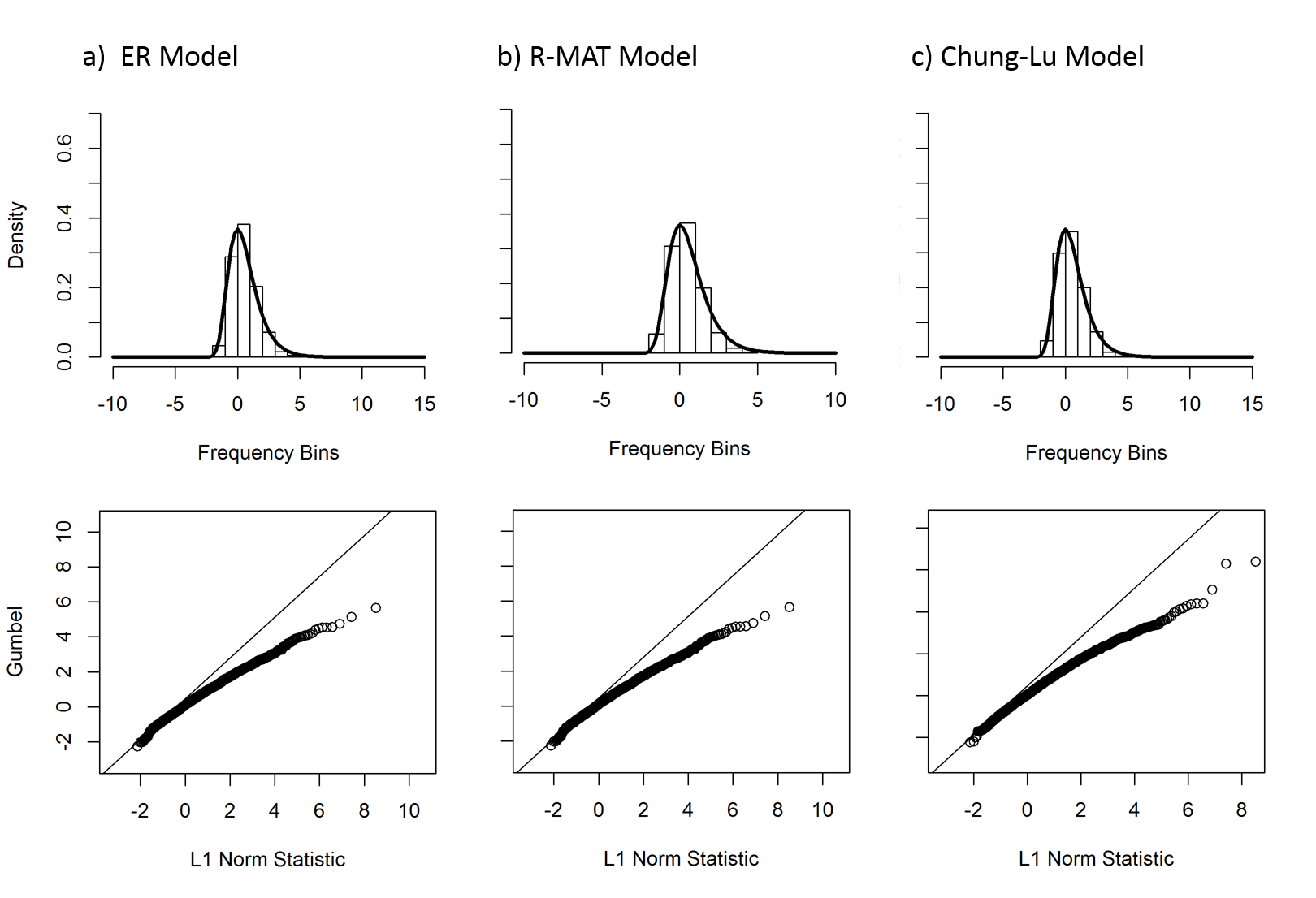}
		\caption{Top figures are histogram density plots when parameters $a_m$ and $b_m$ are estimated using the Extreme Value Theorem with $m < n$. Solid black line represents the theoretical Gumbel distribution. Bottom figures are the corresponding Q-Q plots of the simulated $L_1$ norm statistics with the line $y = x$ representing the theoretical Gumbel distribution and dashed line representing the $99^{th}$ percentile of the theoretical Gumbel distribution. This example is with $n = 512$ and $p_0 = 0.1$; other scenarios are presented in the supplementary material and follow similar patterns. ((a) Erd\"os-R\'enyi, (b) R-MAT, and (c) Chung-Lu Model) }
		\label{fig_HistER_EVL1m512}
	\end{center}
\end{figure*}

Next, we keep $m<n$ and employ the  Extreme Value Theorem to estimate $a_m$ and $b_m$.
  The observations from Figure \ref{fig_HistER_EVL1m512} and Table \ref{tab_10000AllEV} are broadly similar to the MOM results.
  However, the errors are generally higher than using MOM, which is expected since the MOM is based on 1000 historical networks.
  For example in Table \ref{tab_10000AllEV}, the Erd\"os-R\'enyi, R-MAT and Chung-Lu columns corresponding to the 99\% quantile simulation results have wider ranges ($0.74$, $2.55$, and $3.35$) in comparison to Table \ref{tab_10000AllL1m} which are ($0.75$, $0.98$, and $1.49$). 
  For a static network there are no historical data available, so the Extreme Value method can be implemented in applications.
  
\subsubsection{Estimating $a_m$ and $b_m$ using historical data and setting $m = n$}
\begin{figure*}[t]
	\begin{center}
		\includegraphics[height=90mm,width=1.0\linewidth, keepaspectratio=true]{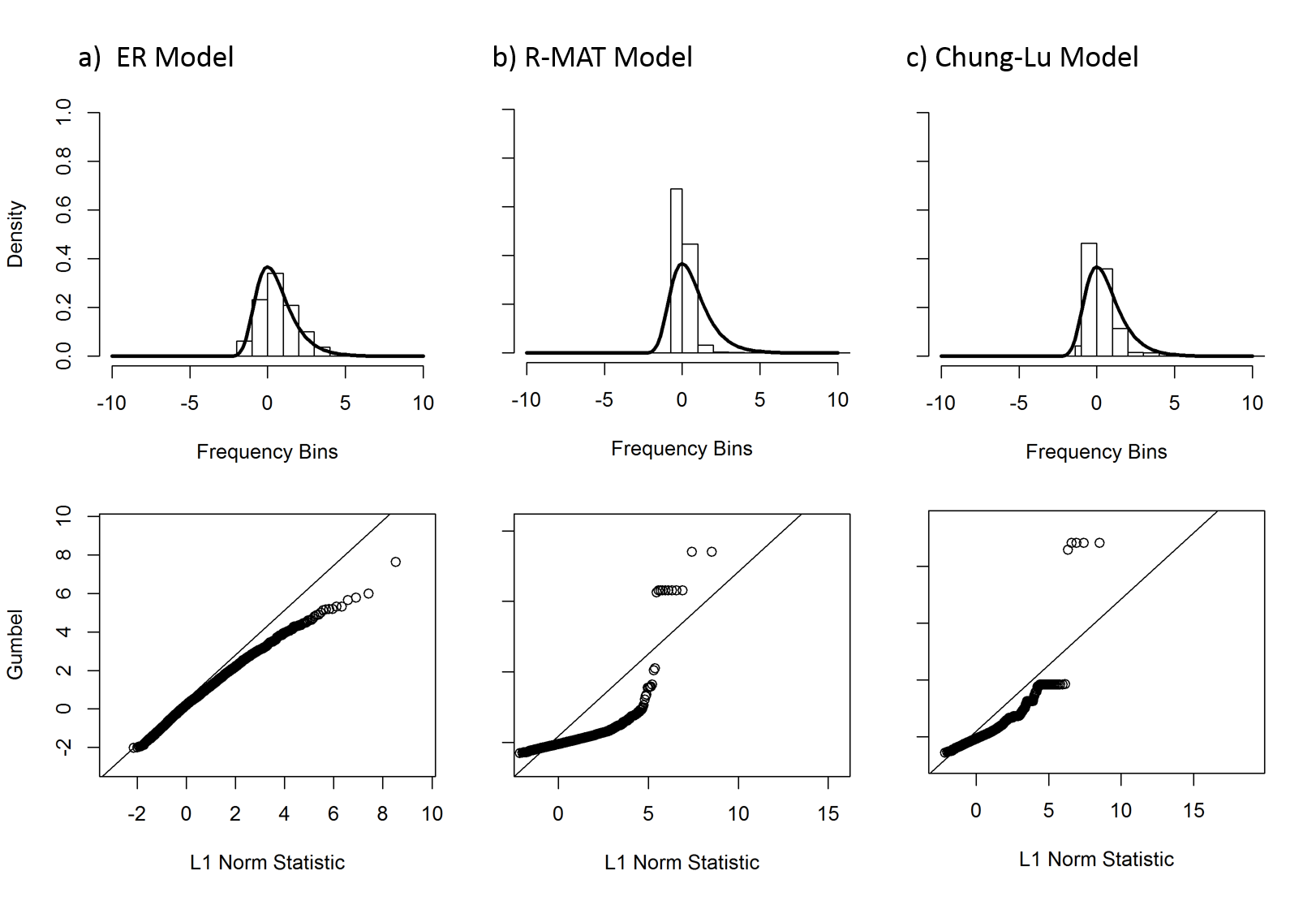}
		\caption{Top figures are histogram density plots when parameters $a_m$ and $b_m$ are estimated using historical data and MOM estimators with $m = n$. Solid black line represents the theoretical Gumbel distribution. Bottom figures are the corresponding Q-Q plots of the simulated $L_1$ norm statistics with the line $y = x$ representing the theoretical Gumbel distribution and dashed line representing the $99^{th}$ percentile of the theoretical Gumbel distribution. This example is with $n = 512$ and $p_0 = 0.1$; other scenarios are presented in the supplementary material and follow similar patterns. ((a) Erd\"os-R\'enyi, (b) R-MAT, and (c) Chung-Lu Model)}
		\label{fig_HistER_L1All512}
	\end{center}
\end{figure*}

We now use $m$ = $n$ with historical data (using MOM estimator) to investigate the impact of changing $m$.
Results are given in Figure \ref{fig_HistER_L1All512} and Table \ref{tab_10000AllL1All}.
Only the ER model yields a distribution similar to the theoretical Gumbel distribution, and the results are generally substantially worse than using $m<n$.
This implies that the performance of the $L_1$ method depends on the type of network model and the number of eigenvectors used. 
It is surprising that using more data ($m=n$ rather than $m<n$) leads to worse performance, since a higher value of $m$ should lead to better fit to the asymptotic Gumbel distribution.
This indicates that in contrast to the assumption in \cite{miller2015spectral}, the null distribution of the $L_1$ statistic is not actually Gumbel.

\subsubsection{Estimating $a_m$ and $b_m$ using the Extreme Value Theorem and setting m = n} 
\begin{figure*}[t]
	\begin{center}
		\includegraphics[height=90mm,width=1.0\linewidth, keepaspectratio=true]{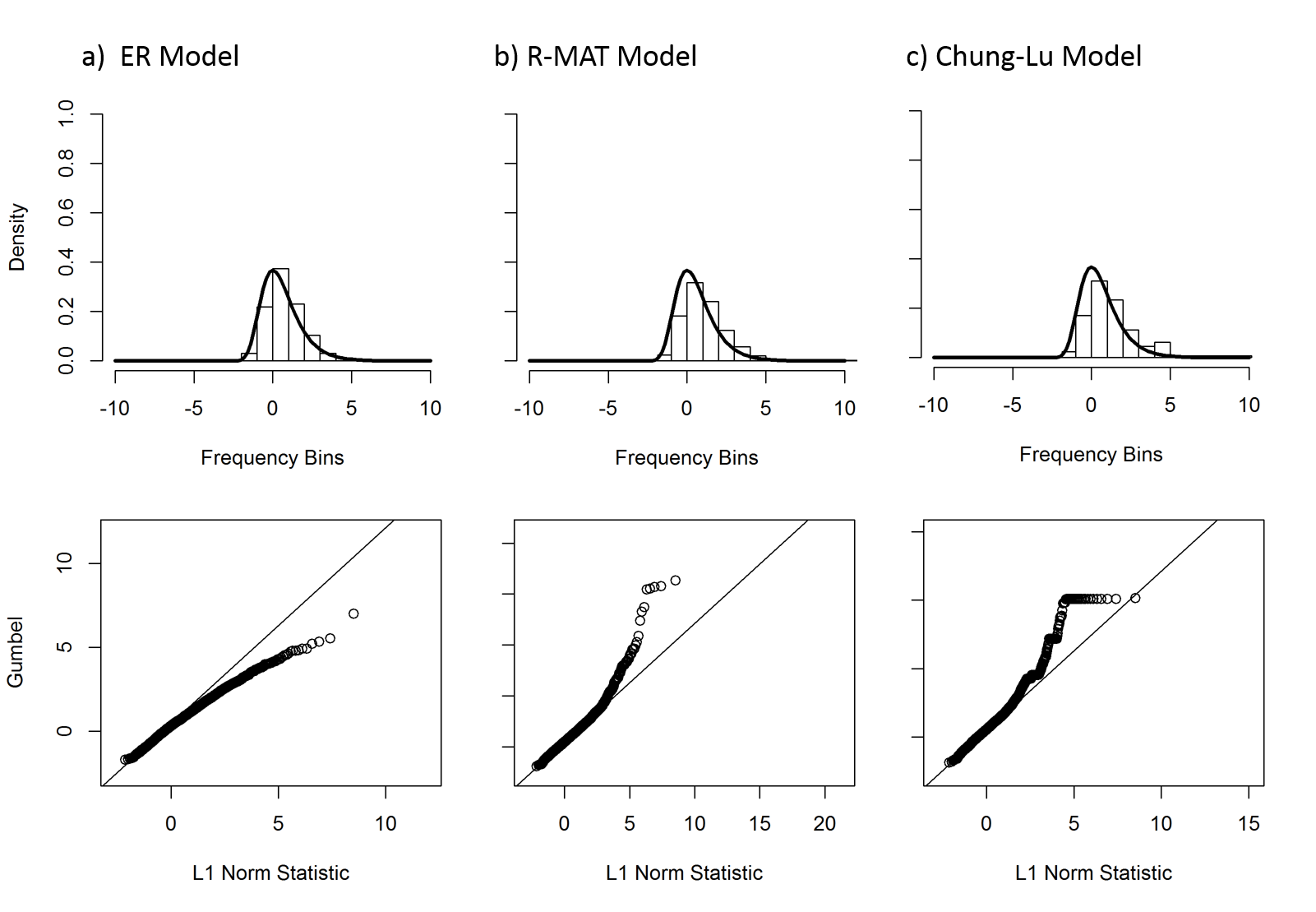}
		\caption{Top figures are histogram density plots when parameters $a_m$ and $b_m$ are estimated using the Extreme Value Theorem with $m = n$. Solid black line represents the theoretical Gumbel distribution. Bottom figures are the corresponding Q-Q plots of the simulated $L_1$ norm statistics with the line $y = x$ representing the theoretical Gumbel distribution and dashed line representing the $99^{th}$ percentile of the theoretical Gumbel distribution. This example is with $n = 512$ and $p_0 = 0.1$; other scenarios are presented in the supplementary material and follow similar patterns. ((a) Erd\"os-R\'enyi, (b) R-MAT, and (c) Chung-Lu Model) }
		\label{fig_HistER_EVAll512}
	\end{center}
\end{figure*}

Finally, we consider $m$ = $n$ in conjunction with estimators using Extreme Value Theorem.
We note that for the three models investigated,  Erd\"os-R\'enyi and R-MAT and Chung-Lu models, the Q-Q plots show varying degrees of differences as seen in Figure \ref{fig_HistER_EVAll512}. For the Erd\"os-R\'enyi model, the histograms and Q-Q plots are similar to the case when $m < n$ using the Extreme Value theorem. However, for the R-MAT and Chung-Lu models, the results are different in comparison to when $m < n$ as illustrated in Figure \ref{fig_HistER_EVAll512} and Table \ref{tab_10000AllEVAll}. 

\subsubsection{Summary of results for the $L_1$ norm algorithm}
The results of a broad range of simulation settings were reported in this section.
It is apparent that the network model, the order of the network and its background connectivity affect the performance of the $L_1$ norm algorithm.  Also, the approach for estimating the parameters has a significant effect on the performance of the algorithms. Using the MOM estimator as compared to using the Extreme Value Theorem for estimating parameters $(a_m, b_m)$ affected performance of the algorithm, particularly for the R-MAT and Chung-Lu models. When using the Extreme Value Theorem, it is expected that the detection statistic should approach the Gumbel distribution as the number of the eigenvectors increases, but we observe the opposite. We showed that the distribution of the detection statistic does not follow the Gumbel distribution with particularly large deviations at the higher quantiles. None of the proposed methods worked well for all models investigated. 

A possible explanation for the $L_1$ norm technique performing worse with more eigenvectors in the $m=n$ case is sparsity.
Sparse networks are likely to be disconnected, leading to one or more small eigenvalues.
The eigenvectors associated with these small eigenvalues are likely to be very unstructured, leading to high variability and non-normality.

\begin{table*}[h]
\caption{Quantiles of the $L_1$ norm based on 10,000 simulations. No anomalous subgraph is present. The results are compared to the theoretical Gumbel distribution with $m$ = 30 for $n = 128$ and $n= 256$, and $m$ = 50 for $n = 512$ and $n=1024$. Scaling parameters $a_m$ and $b_m$ are estimated from historical data using MOM estimators}
\label{tab_10000AllL1m}
\begin{adjustbox}{width=1\textwidth}
\centering
\begin{tabular}{|rr|rrrrr|rrrrr|rrrrr|}
\hline
\multicolumn{2}{|l|} {} &  \multicolumn{5}{|c|}{ER Model} & \multicolumn{5}{|c|}{R-MAT Model} &  \multicolumn{5}{|c|}{Chung-Lu Model} \\  \hline  
 Network order & $p_0$ & $95\%$ & $96\%$ & $97\%$ & $98\%$ & $99\%$ & $95\%$ & $96\%$ & $97\%$ & $98\%$ & $99\%$& $95\%$ & $96\%$ & $97\%$ & $98\%$ & $99\%$\\ 
  \hline
128 & 0.050 & 2.94 & 3.14 & 3.38 & 3.71 & 4.23 & 2.89 & 3.09 & 3.38 & 3.95 & 4.75 & 2.70 & 2.99 & 3.26 & 3.75 & 5.28 \\ 
  128 & 0.100 & 2.87 & 3.07 & 3.40 & 3.68 & 4.23 & 2.78 & 2.97 & 3.24 & 3.65 & 4.17 & 3.01 & 3.24 & 3.46 & 3.77 & 4.37 \\ 
  128 & 0.300 & 2.75 & 2.90 & 3.08 & 3.43 & 3.94 & 2.62 & 2.80 & 3.10 & 3.34 & 3.77 & 2.74 & 2.92 & 3.14 & 3.45 & 3.80 \\ 
  256 & 0.010 & 2.79 & 2.91 & 3.16 & 3.63 & 4.18 & 2.97 & 3.19 & 3.44 & 3.76 & 4.15 & 2.85 & 3.02 & 3.33 & 3.70 & 4.08 \\ 
  256 & 0.100 & 3.05 & 3.26 & 3.47 & 3.84 & 4.38 & 2.86 & 3.08 & 3.35 & 3.67 & 4.21 & 2.69 & 2.91 & 3.10 & 3.43 & 3.80 \\ 
  256 & 0.300 & 2.89 & 3.11 & 3.34 & 3.65 & 4.19 & 3.00 & 3.15 & 3.44 & 3.76 & 4.13 & 2.98 & 3.18 & 3.40 & 3.67 & 4.37 \\ 
  512 & 0.010 & 2.79 & 2.93 & 3.20 & 3.48 & 4.07 & 2.83 & 3.05 & 3.26 & 3.60 & 4.06 & 2.70 & 2.81 & 3.00 & 3.32 & 3.86 \\ 
  512 & 0.100 & 3.09 & 3.33 & 3.53 & 3.91 & 4.69 & 3.06 & 3.21 & 3.45 & 3.86 & 4.33 & 3.04 & 3.21 & 3.48 & 3.74 & 4.21 \\ 
  512 & 0.300 & 3.09 & 3.23 & 3.42 & 3.64 & 4.01 & 2.89 & 3.00 & 3.23 & 3.56 & 4.05 & 2.99 & 3.18 & 3.47 & 3.71 & 4.25 \\ 
  1024 & 0.010 & 2.77 & 2.97 & 3.24 & 3.60 & 4.06 & 2.94 & 3.14 & 3.36 & 3.66 & 4.34 & 2.91 & 3.20 & 3.45 & 3.90 & 4.57 \\ 
  1024 & 0.100 & 3.09 & 3.28 & 3.51 & 3.74 & 4.42 & 2.90 & 3.04 & 3.27 & 3.62 & 4.10 & 2.81 & 2.98 & 3.25 & 3.57 & 4.07 \\ 
  1024 & 0.300 & 3.04 & 3.25 & 3.52 & 3.81 & 4.27 & 3.02 & 3.21 & 3.59 & 3.99 & 4.41 & 2.97 & 3.20 & 3.46 & 3.69 & 4.10 \\ 
   \hline
    \multicolumn{2}{|c|} {\textbf{Gumbel quantiles}} & \textbf{2.97} & \textbf{3.20} & \textbf{3.49} & \textbf{3.90} & \textbf{4.60} & \textbf{2.97} & \textbf{3.20} & \textbf{3.49} & \textbf{3.90} & \textbf{4.60} & \textbf{2.97} & \textbf{3.20} & \textbf{3.49} & \textbf{3.90} & \textbf{4.60} \\ \hline
\end{tabular}
\end{adjustbox}
\end{table*}
\clearpage

\begin{table*}[b]
\caption{Quantiles of the $L_1$ norm based on 10,000 simulations. No anomalous subgraph is present. The results are compared to the theoretical Gumbel distribution with $m$ = 30 for $n = 128$ and $n= 256$, and $m$ = 50 for $n = 512$ and $n= 1024$. Scaling parameters $a_m$ and $b_m$ are estimated using the Extreme Value Theorem}
\label{tab_10000AllEV}
\begin{adjustbox}{width=1\textwidth}
\centering
\begin{tabular}{|rr|rrrrr|rrrrr|rrrrr|}
\hline
\multicolumn{2}{|l|} {} &  \multicolumn{5}{|c|}{ER Model} & \multicolumn{5}{|c|}{R-MAT Model} &  \multicolumn{5}{|c|}{Chung-Lu Model} \\  \hline  
 Network order & $p_0$ & $95\%$ & $96\%$ & $97\%$ & $98\%$ & $99\%$ & $95\%$ & $96\%$ & $97\%$ & $98\%$ & $99\%$& $95\%$ & $96\%$ & $97\%$ & $98\%$ & $99\%$\\ 
  \hline
128 & 0.050 & 2.71 & 2.89 & 3.10 & 3.39 & 3.85 & 3.35 & 3.57 & 3.90 & 4.56 & 5.47 & 3.20 & 3.55 & 3.87 & 4.40 & 6.22 \\ 
  128 & 0.100 & 2.47 & 2.63 & 2.89 & 3.12 & 3.57 & 2.42 & 2.58 & 2.80 & 3.15 & 3.59 & 2.52 & 2.71 & 2.89 & 3.15 & 3.65 \\ 
  128 & 0.300 & 2.27 & 2.39 & 2.53 & 2.81 & 3.21 & 2.10 & 2.24 & 2.48 & 2.67 & 3.02 & 2.20 & 2.34 & 2.52 & 2.76 & 3.04 \\ 
  256 & 0.010 & 2.52 & 2.63 & 2.85 & 3.25 & 3.73 & 2.37 & 2.54 & 2.74 & 3.00 & 3.31 & 2.38 & 2.52 & 2.79 & 3.09 & 3.41 \\ 
  256 & 0.100 & 2.50 & 2.67 & 2.83 & 3.13 & 3.56 & 2.16 & 2.33 & 2.54 & 2.78 & 3.19 & 2.13 & 2.30 & 2.44 & 2.70 & 2.99 \\ 
  256 & 0.300 & 2.22 & 2.38 & 2.56 & 2.78 & 3.19 & 2.07 & 2.19 & 2.41 & 2.64 & 2.92 & 2.22 & 2.36 & 2.52 & 2.73 & 3.24 \\ 
  512 & 0.010 & 2.49 & 2.61 & 2.84 & 3.07 & 3.57 & 2.34 & 2.51 & 2.67 & 2.94 & 3.31 & 2.37 & 2.47 & 2.63 & 2.91 & 3.38 \\ 
  512 & 0.100 & 2.41 & 2.59 & 2.74 & 3.03 & 3.63 & 2.32 & 2.43 & 2.62 & 2.93 & 3.30 & 2.40 & 2.54 & 2.74 & 2.94 & 3.30 \\ 
  512 & 0.300 & 2.41 & 2.52 & 2.66 & 2.83 & 3.11 & 2.19 & 2.28 & 2.46 & 2.72 & 3.11 & 2.38 & 2.53 & 2.76 & 2.95 & 3.37 \\ 
  1024 & 0.010 & 2.37 & 2.54 & 2.77 & 3.07 & 3.45 & 2.28 & 2.44 & 2.61 & 2.85 & 3.39 & 2.38 & 2.61 & 2.81 & 3.17 & 3.71 \\ 
  1024 & 0.100 & 2.38 & 2.53 & 2.69 & 2.87 & 3.37 & 2.41 & 2.53 & 2.72 & 3.02 & 3.42 & 2.25 & 2.39 & 2.59 & 2.84 & 3.23 \\ 
  1024 & 0.300 & 2.33 & 2.49 & 2.70 & 2.92 & 3.28 & 2.19 & 2.33 & 2.62 & 2.93 & 3.25 & 2.32 & 2.50 & 2.71 & 2.88 & 3.21 \\ 
   \hline
   \multicolumn{2}{|c|} {\textbf{Gumbel quantiles}} & \textbf{2.97} & \textbf{3.20} & \textbf{3.49} & \textbf{3.90} & \textbf{4.60} & \textbf{2.97} & \textbf{3.20} & \textbf{3.49} & \textbf{3.90} & \textbf{4.60} & \textbf{2.97} & \textbf{3.20} & \textbf{3.49} & \textbf{3.90} & \textbf{4.60} \\ \hline
\end{tabular}
\end{adjustbox}
\end{table*}
\clearpage

\setlength{\floatsep}{5pt plus 1.0pt minus 2.0pt}

\begin{table*}[b]
\caption{Quantiles of the $L_1$ norm based on 10,000 simulations. No anomalous subgraph is present. The results are compared to the theoretical Gumbel distribution with $m = n$. Scaling parameters $a_m$ and $b_m$ are estimated from historical data using MOM estimators}
\label{tab_10000AllL1All}
\begin{adjustbox}{width=1\textwidth}
\centering
\begin{tabular}{|rr|rrrrr|rrrrr|rrrrr|}
\hline
\multicolumn{2}{|l|} {} &  \multicolumn{5}{|c|}{ER Model} & \multicolumn{5}{|c|}{R-MAT Model} &  \multicolumn{5}{|c|}{Chung-Lu Model} \\  \hline  
 Network order & $p_0$ & $95\%$ & $96\%$ & $97\%$ & $98\%$ & $99\%$ & $95\%$ & $96\%$ & $97\%$ & $98\%$ & $99\%$& $95\%$ & $96\%$ & $97\%$ & $98\%$ & $99\%$\\ 
  \hline
128 & 0.050 & 1.63 & 1.98 & 2.49 & 6.18 & 7.33 & 3.35 & 3.75 & 3.81 & 5.18 & 6.31 & 2.37 & 2.69 & 3.39 & 3.97 & 5.70 \\ 
  128 & 0.100 & 3.21 & 3.34 & 3.52 & 3.89 & 4.62 & 2.33 & 2.94 & 2.99 & 3.43 & 7.49 & 2.11 & 2.25 & 2.39 & 2.53 & 7.14 \\ 
  128 & 0.300 & 2.93 & 3.19 & 3.48 & 3.73 & 4.23 & 2.80 & 3.05 & 3.25 & 3.62 & 4.24 & 1.75 & 1.88 & 2.14 & 2.46 & 3.03 \\ 
  256 & 0.010 & 3.15 & 3.34 & 3.60 & 4.07 & 4.56 & 2.33 & 3.50 & 4.42 & 5.03 & 7.81 & 2.11 & 3.87 & 4.64 & 5.87 & 9.54 \\ 
  256 & 0.100 & 3.07 & 3.30 & 3.66 & 3.96 & 4.65 & 1.25 & 2.92 & 2.92 & 4.38 & 5.32 & 1.86 & 2.11 & 2.13 & 3.04 & 3.22 \\ 
  256 & 0.300 & 3.24 & 3.40 & 3.67 & 3.95 & 4.71 & 3.17 & 3.50 & 3.76 & 4.08 & 4.61 & 2.71 & 3.00 & 3.25 & 3.70 & 4.52 \\ 
  512 & 0.010 & 3.14 & 3.32 & 3.59 & 3.81 & 4.27 & 1.31 & 1.68 & 1.79 & 3.55 & 4.11 & 1.17 & 1.45 & 2.14 & 4.10 & 5.02 \\ 
  512 & 0.100 & 3.07 & 3.22 & 3.50 & 3.85 & 4.35 & 1.00 & 1.20 & 1.50 & 1.95 & 10.66 & 1.91 & 2.35 & 3.13 & 3.49 & 4.59 \\ 
  512 & 0.300 & 3.10 & 3.33 & 3.54 & 3.98 & 4.56 & 3.05 & 3.33 & 3.61 & 4.13 & 4.70 & 3.03 & 3.19 & 3.50 & 3.86 & 4.56 \\ 
  1024 & 0.010 & 3.15 & 3.41 & 3.61 & 4.04 & 4.76 & 1.93 & 2.15 & 2.15 & 2.27 & 10.07 & 2.17 & 3.53 & 4.80 & 5.50 & 5.66 \\ 
  1024 & 0.100 & 3.26 & 3.46 & 3.78 & 4.17 & 4.63 & 2.93 & 3.15 & 3.38 & 3.92 & 4.82 & 1.78 & 2.15 & 3.35 & 4.11 & 4.78 \\ 
  1024 & 0.300 & 3.30 & 3.48 & 3.85 & 4.14 & 4.66 & 3.30 & 3.49 & 3.70 & 4.06 & 4.59 & 3.11 & 3.29 & 3.58 & 3.90 & 4.51 \\ 
   \hline
    \multicolumn{2}{|c|} {\textbf{Gumbel quantiles}} & \textbf{2.97} & \textbf{3.20} & \textbf{3.49} & \textbf{3.90} & \textbf{4.60} & \textbf{2.97} & \textbf{3.20} & \textbf{3.49} & \textbf{3.90} & \textbf{4.60} & \textbf{2.97} & \textbf{3.20} & \textbf{3.49} & \textbf{3.90} & \textbf{4.60} \\ \hline
\end{tabular}
\end{adjustbox}
\end{table*}
\clearpage

\begin{table*}[b]
\caption{Quantiles of the $L_1$ norm based on 10,000 simulations. No anomalous subgraph is present. The results are compared to the theoretical Gumbel distribution with $m = n$. Scaling parameters $a_m$ and $b_m$ are estimated using the Extreme Value Theorem}
\label{tab_10000AllEVAll}
\begin{adjustbox}{width=1\textwidth}
\centering
\begin{tabular}{|rr|rrrrr|rrrrr|rrrrr|}
\hline
\multicolumn{2}{|l|} {} &  \multicolumn{5}{|c|}{ER Model} & \multicolumn{5}{|c|}{R-MAT Model} &  \multicolumn{5}{|c|}{Chung-Lu Model} \\  \hline  
 Network order & $p_0$ & $95\%$ & $96\%$ & $97\%$ & $98\%$ & $99\%$ & $95\%$ & $96\%$ & $97\%$ & $98\%$ & $99\%$& $95\%$ & $96\%$ & $97\%$ & $98\%$ & $99\%$\\ 
 \hline
128 & 0.050 & 5.01 & 5.87 & 7.15 & 16.53 & 19.44 & 18.79 & 20.68 & 20.94 & 27.37 & 32.66 & 10.78 & 12.05 & 14.81 & 17.11 & 23.94 \\ 
  128 & 0.100 & 3.37 & 3.50 & 3.67 & 4.03 & 4.73 & 10.14 & 11.99 & 13.19 & 13.32 & 19.90 & 7.74 & 8.16 & 8.63 & 9.00 & 21.92 \\ 
  128 & 0.300 & 2.82 & 3.05 & 3.31 & 3.53 & 3.98 & 3.01 & 3.26 & 3.46 & 3.82 & 4.43 & 3.25 & 3.46 & 3.89 & 4.38 & 5.25 \\ 
  256 & 0.010 & 3.41 & 3.61 & 3.86 & 4.33 & 4.81 & 12.16 & 16.38 & 22.66 & 27.89 & 31.34 & 7.75 & 10.64 & 17.56 & 20.40 & 30.02 \\ 
  256 & 0.100 & 2.92 & 3.13 & 3.46 & 3.73 & 4.35 & 6.65 & 11.89 & 14.57 & 19.33 & 20.97 & 6.24 & 7.17 & 7.25 & 9.82 & 10.41 \\ 
  256 & 0.300 & 2.94 & 3.08 & 3.32 & 3.56 & 4.23 & 3.04 & 3.34 & 3.57 & 3.86 & 4.34 & 3.37 & 3.70 & 3.98 & 4.49 & 5.41 \\ 
  512 & 0.010 & 3.08 & 3.25 & 3.50 & 3.70 & 4.13 & 7.91 & 10.93 & 11.57 & 13.95 & 22.76 & 4.88 & 5.70 & 7.75 & 14.46 & 17.43 \\ 
  512 & 0.100 & 2.90 & 3.04 & 3.29 & 3.60 & 4.06 & 4.41 & 5.02 & 5.66 & 6.89 & 9.26 & 4.73 & 5.62 & 7.19 & 7.90 & 10.10 \\ 
  512 & 0.300 & 2.89 & 3.09 & 3.27 & 3.67 & 4.17 & 2.96 & 3.22 & 3.47 & 3.94 & 4.46 & 3.10 & 3.26 & 3.55 & 3.89 & 4.56 \\ 
  1024 & 0.010 & 2.98 & 3.22 & 3.41 & 3.80 & 4.46 & 11.95 & 11.95 & 13.13 & 13.13 & 13.76 & 5.42 & 7.93 & 11.58 & 11.79 & 13.57 \\ 
  1024 & 0.100 & 3.03 & 3.21 & 3.49 & 3.84 & 4.25 & 3.49 & 3.73 & 3.98 & 4.58 & 5.57 & 5.31 & 6.11 & 8.38 & 11.17 & 12.83 \\ 
  1024 & 0.300 & 2.96 & 3.12 & 3.46 & 3.71 & 4.18 & 2.99 & 3.16 & 3.35 & 3.66 & 4.13 & 3.10 & 3.26 & 3.53 & 3.84 & 4.41 \\ 
   \hline
   \multicolumn{2}{|c|} {\textbf{Gumbel quantiles}} & \textbf{2.97} & \textbf{3.20} & \textbf{3.49} & \textbf{3.90} & \textbf{4.60} & \textbf{2.97} & \textbf{3.20} & \textbf{3.49} & \textbf{3.90} & \textbf{4.60} & \textbf{2.97} & \textbf{3.20} & \textbf{3.49} & \textbf{3.90} & \textbf{4.60} \\ \hline
\end{tabular}
\end{adjustbox}
\end{table*}
\clearpage

\subsection{Improving the $L_1$ norm algorithm}
As mentioned in previous sections, two main concerns arise when implementing the $L_1$ norm algorithm:

\begin{enumerate}
       \item The need for historical data for estimating the detection statistic parameters.
       \item The number of eigenvectors, $m$, to use from the eigenspace.
\end{enumerate}
In \cite{miller2015spectral}, there is no discussion on how the Gumbel distribution parameters, $a_m$ and $b_m$, as well as the detection statistic parameters, mean $\mu_k$ and standard deviation $\sigma_k$, should be estimated, especially as historical data are needed. However, we proposed estimating the location and scaling parameter, $a_m$ and $b_m$, using the Extreme Value Theorem as it does not require historical data. To estimate the detection statistic parameters,  $\mu_k$ and  $\sigma_k$, we developed an approach that only requires the current static network. 
The $L_1$ norm proposed by \cite{miller2015spectral} is
\begin{equation} \label{eq_L1min}
L = -\min_{1 \leq k \leq m}  \frac{|\textbf{X}_k|_1 - \mu_k}{\sigma_k},
\end{equation}
where $\mu_k$ and $\sigma_k$ are the mean and standard deviation of the $k^{th}$ eigenvector of the residual matrix, estimated using historical data with no anomalies. 
Again, the implementation of this algorithm is impractical for most static networks since historical data are needed. To overcome this issue, we analyze how this statistic performs if $|\textbf{X}_k|_1$ is standardized using only the $m$ eigenvectors of the current network. We studied three different standardization approaches as follows.
The constants $k_1$ and $k_2$ follow from the relation between standard deviation, interquartile range, and mean absolute deviation of the normal distribution.
\begin{enumerate}
	\item Using mean and standard deviation of ($m$) $L_1$ norms of the adjacency matrix:
	$$L = -\min_{1 \leq k \leq m}  \frac{|\textbf{X}_k|_1 - \hat\mu}{\hat\sigma}.$$
	\item Using median and IQR (Inter Quartile Range) of ($m$) $L_1$ norms of the adjacency matrix:
	$$L = -\min_{1 \leq k \leq m}  \frac{|\textbf{X}_k|_1 - M}{IQR/k_1}.$$
	where $M$ is the median of $m$ $L_1$ norms for the current network, and $k_1=1.3489$, assuming that the eigenvectors of the residual matrix follow a normal distribution.
	\item Using median and mad (median absolute deviation) of $m$ $L_1$ norms of the adjacency matrix:
	$$L = -\min_{1 \leq k \leq m}  \frac{|\textbf{X}_k|_1 - M}{mad/k_2}$$
	where $M$ is the median of $m$ $L_1$ norms for the current network, and $k_2=0.67449$, assuming that the $L_1$ norms of the eigenvectors follow a normal distribution.
\end{enumerate}

Using the median and IQR performed the best.  This is because, if an anomalous subgraph is present, the median, $M$, and interquartile range $IQR$, will not be affected which makes this approach appropriate for standardizing the detection statistic. Also, as in previous explorations, see Table \ref{tab_10000AllEV} and Table \ref{tab_10000AllEVAll}, selecting $m < n$ also worked the best. In particular, using an $m$ between 30 to 50 provides the best results in most of the network combinations we explored where $m = 30$ applies to smaller networks ($n < 257$) and $m = 50$ is suggested for larger networks ($1025 > n > 257$). Finally, we see that approximating the $L_1$ norm statistic using the eigenvectors of a single network performs sufficiently well. Quantiles corresponding to the best performing methodology, i.e., using the median and IQR, are shown in Figure \ref{fig_L1_IQR} and Table \ref{tab_10000med_IQR} below. We have provided a comprehensive review of our simulation results as well as detection and false alarm rate performance for all three alternatives in the supplementary material.

\begin{figure*}[h]
	\begin{center}
		\includegraphics[height=0.40 \textheight, keepaspectratio=true]{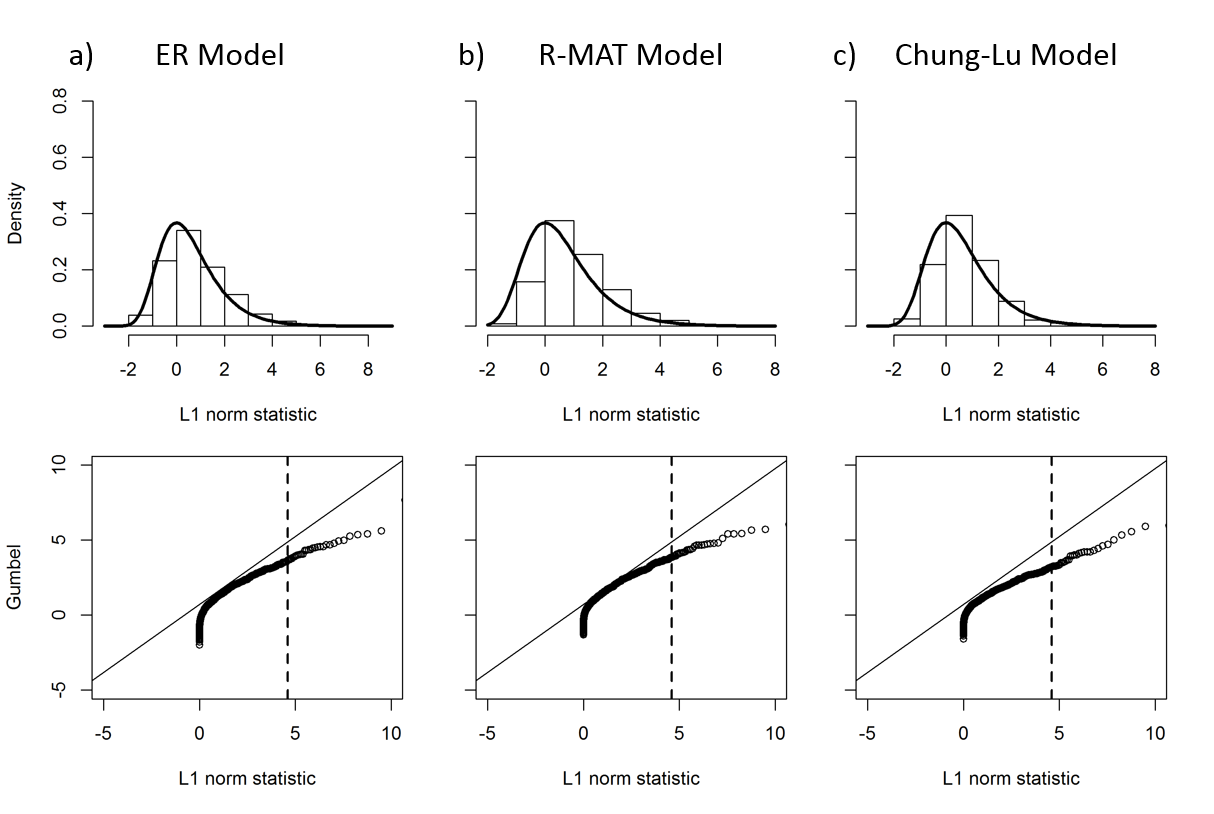}
		\caption{Top figures are histogram density plots based on 10,000 simulations using inter-quantile range, $IQR$, and the median, $M$ to standardize the $L_1$ norm detection statistic. $m = 50$, $n = 512$, $p_0 = 0.1$.  Bottom figures are the corresponding Q-Q plots with the dashed line representing the $99^{th}$ percentile of the theoretical Gumbel distribution. ((a) Erd\"os-R\'enyi, (b) R-MAT, and (c) Chung-Lu Model)}
		\label{fig_L1_IQR}
	\end{center}
\end{figure*}

\begin{table*}[h]
\caption{($L_1$ norm, $m < n$, Median and IQR) 10,000 in-control simulations are run and the results compared to the theoretical Gumbel distribution when $m = 30$ for $n = 128$ and $n=256$, and $m = 50$ for $n = 512$ and $n=1024$.}
\label{tab_10000med_IQR}
\begin{adjustbox}{width=1\textwidth}
\centering
\begin{tabular}{|rr|rrrrr|rrrrr|rrrrr|}
\hline
\multicolumn{2}{|l|} {} &  \multicolumn{5}{|c|}{ER Model} & \multicolumn{5}{|c|}{R-MAT Model} &  \multicolumn{5}{|c|}{Chung-Lu Model} \\  \hline  
 Network order & $p_0$ & $95\%$ & $96\%$ & $97\%$ & $98\%$ & $99\%$ & $95\%$ & $96\%$ & $97\%$ & $98\%$ & $99\%$& $95\%$ & $96\%$ & $97\%$ & $98\%$ & $99\%$\\ 
  \hline

  128 & 0.050 & 6.55 & 6.90 & 7.52 & 8.56 & 10.09 & 3.81 & 4.05 & 4.45 & 4.71 & 5.14 & 4.25 & 4.49 & 4.92 & 5.39 & 6.36 \\ 
  128 & 0.100 & 4.30 & 4.50 & 5.03 & 5.86 & 6.74 & 3.16 & 3.39 & 3.80 & 4.01 & 4.49 & 3.58 & 3.78 & 4.07 & 4.27 & 5.17 \\ 
  128 & 0.300 & 3.26 & 3.48 & 3.88 & 4.26 & 5.31 & 3.51 & 3.70 & 4.02 & 4.46 & 5.40 & 3.35 & 3.52 & 3.88 & 4.26 & 5.03 \\ 
  256 & 0.010 & 7.37 & 7.60 & 8.28 & 8.79 & 9.77 & 4.26 & 4.41 & 4.71 & 5.13 & 6.14 & 4.69 & 4.96 & 5.30 & 5.82 & 6.62 \\ 
  256 & 0.100 & 4.13 & 4.35 & 4.65 & 5.06 & 6.11 & 2.41 & 2.56 & 2.77 & 3.03 & 3.86 & 3.35 & 3.53 & 3.76 & 4.12 & 4.88 \\ 
  256 & 0.300 & 3.53 & 3.87 & 4.34 & 4.90 & 5.87 & 3.31 & 3.43 & 3.79 & 4.33 & 4.77 & 3.60 & 4.01 & 4.26 & 4.72 & 5.44 \\ 
  512 & 0.010 & 10.53 & 10.73 & 11.53 & 12.18 & 13.16 & 3.14 & 3.27 & 3.41 & 3.59 & 3.83 & 5.43 & 5.69 & 5.94 & 6.52 & 7.02 \\ 
  512 & 0.100 & 3.17 & 3.39 & 3.68 & 4.08 & 4.68 & 3.50 & 3.68 & 3.93 & 4.35 & 4.76 & 2.77 & 2.98 & 3.26 & 3.66 & 4.22 \\ 
  512 & 0.300 & 3.18 & 3.47 & 3.79 & 4.16 & 4.87 & 4.24 & 4.57 & 4.71 & 5.01 & 5.76 & 3.14 & 3.27 & 3.51 & 4.08 & 4.74 \\ 
  1024 & 0.010 & 8.91 & 9.70 & 10.21 & 11.15 & 13.36 & 1.48 & 1.58 & 1.68 & 1.82 & 2.01 & 3.85 & 3.98 & 4.22 & 4.60 & 5.02 \\ 
  1024 & 0.100 & 3.44 & 3.65 & 3.92 & 4.36 & 5.09 & 8.05 & 8.81 & 9.42 & 9.99 & 10.71 & 2.45 & 2.60 & 2.81 & 3.36 & 3.84 \\ 
  1024 & 0.300 & 3.29 & 3.58 & 3.83 & 4.30 & 4.78 & 7.73 & 7.98 & 8.19 & 8.89 & 9.43 & 3.15 & 3.41 & 3.67 & 4.06 & 4.56 \\ 
   \hline
    \multicolumn{2}{|c|} {\textbf{Gumbel quantiles}} & \textbf{2.97} & \textbf{3.20} & \textbf{3.49} & \textbf{3.90} & \textbf{4.60} & \textbf{2.97} & \textbf{3.20} & \textbf{3.49} & \textbf{3.90} & \textbf{4.60} & \textbf{2.97} & \textbf{3.20} & \textbf{3.49} & \textbf{3.90} & \textbf{4.60} \\ \hline
\end{tabular}
\end{adjustbox}
\end{table*}

\section{Evaluating algorithm performance}


In this section, we evaluate both algorithms proposed by \cite{miller2015spectral} for the case when an anomaly is present and for different network orders and background connectivity. Following the discussion on the $L_1$ norm algorithm in Section 4, we note that the use of historical data, as suggested in \cite{miller2015spectral} is not a feasible option for static networks.
Therefore, we do not consider the MOM estimator, and focus on the $m<n$ and $m=n$ cases using the Extreme Value Theorem estimators for standardization.
Thus, we focus on three anomaly detection methods for performance evaluation --- the $\chi^2$ algorithm, the $L_1$ norm algorithm with $m<n$, and the $L_1$ norm algorithm with $m=n$.
Table \ref{tab_EROoC} illustrates our results. 

For the chi-square algorithm, in all the cases explored, the false alarm rate from using the chi-square distribution is significantly higher than the expected false alarm rate of $0.05$. Although the detection rate is high, having significantly higher false alarm rates than expected results in an algorithm that is impractical to implement in practice. This again highlights that the chi-square distribution does not provide the appropriate detection threshold for use in anomaly detection. Instead, some method for improving the algorithm is needed. 
We observe the same scenario for the R-MAT model in Table \ref{tab_EROoC}. For the $\chi^2$ value of $3.84$ corresponding to the 95\% theoretical chi-square percentile with $df$ = 1, the false alarm rates are inconsistent for different network order and background probability combinations. This emphasizes again that the algorithm statistic detection threshold selected is dependent on the network model being investigated. This is also true for the Chung-Lu model as shown in Table \ref{tab_EROoC}. That is, the chi-square detection value for $\alpha = 0.05$ produces a false alarm rate (FAR) that exceeds the desired FAR rate of 5\% in all cases.  

On the other hand, the $L_1$ norm algorithm performs much better with respect to detection and false alarm rates. For all networks observed, the $L_1$ norm algorithm has false alarm rates that are relatively close to the desired false alarm rate. Furthermore, the detection rates are relatively high for all network orders and connectivities investigated.

\begin{table*}[h] 
      \centering
      \caption{Comparison of detection and False Alarm Rates for the three explored cases,  $\chi^2$ algorithm, $L_1$ norm using Extreme Value Theorem and $m<n$,  $L_1$ norm using Extreme Value Theorem and $m=n$. Background probability is $p_0 = 0.01$ and foreground probability, with clique present, is $p_1 = 1$. We performed 500 simulations for each network order and connectivity combination with an anomalous subgraph randomly embedded in 250 of 500 simulations. $95^{th}$ percentile was used for signaling threshold}
\label{tab_EROoC}
\begin{adjustbox}{width=1\textwidth}
\small\begin{tabular}{|c|l|l|l|l|l|l|l|}
  \hline
 \multicolumn{2}{l} {\textbf{Erd\"os-R\'enyi Model}} &  \multicolumn{3}{|c|}{Detection Rate \%} & \multicolumn{3}{|c|}{False Alarm Rate \%}\\  \hline  
Network order  & Subgraph Size & $\chi^2$ Ther.  & $L_1$ EV $m < n$  &  $L_1$ EV $m = n$  &  $\chi^2$ Ther.  & $L_1$ EV $m < n$ & $L_1$ EV $m = n$ \\ \hline
  \hline
256 & 8 & 100.00 & 100.00 & 100.00 & 83.70 & 9.90 & 5.50 \\ 
  256 & 10 & 100.00 & 100.00 & 100.00 & 83.70 &  9.90  & 5.50 \\ 
  256 & 13 & 100.00 & 100.00 & 100.00 & 83.70 &  9.90  & 5.50 \\ 
  256 & 15 & 100.00 & 100.00 & 100.00 & 83.70 &  9.90  & 5.50 \\ 
   \hline
  \hline
512& 5 & 100.00 & 100.00 & 100.00 & 31.00 & 5.50 & 2.60 \\ 
  512 & 10 & 100.00 & 100.00 & 100.00 & 31.00  & 5.50 & 2.60 \\ 
  512 & 15 & 100.00 & 100.00 & 100.00 & 31.00  & 5.50 & 2.60 \\ 
  512 & 20 & 100.00 & 100.00 & 100.00 & 31.00  & 5.50 & 2.60 \\ 
   \hline
   \multicolumn{2}{l} {\textbf{R-MAT Model}} &  \multicolumn{3}{|c|}{Detection Rate} & \multicolumn{3}{|c|}{False Alarm Rate}\\  \hline \hline
256 & 8 & 100.00 & 100.00 & 100.00 & 99.30 & 15.60 & 11.40 \\ 
  256 & 10 & 100.00 & 100.00 & 100.00 & 99.30 & 15.60 & 11.40 \\ 
  256 & 13 & 100.00 & 100.00 & 100.00 & 99.30 & 15.60 & 11.40 \\ 
  256 & 15 & 100.00 & 100.00 & 100.00 & 99.30 & 15.60 & 11.40 \\ 
   \hline
512 & 5 & 94.40 & 20.40 & 13.20 & 92.80 & 4.90 & 1.70 \\ 
  512 & 10 & 100.00 & 100.00 & 100.00 & 92.80 & 4.90 & 1.70 \\ 
  512 & 15 & 100.00 & 100.00 & 100.00 & 92.80 & 4.90 & 1.70 \\ 
  512 & 20 & 100.00 & 100.00 & 100.00 & 92.80 & 4.90 & 1.70 \\  \hline
 \multicolumn{2}{l} {\textbf{Chung-Lu Model}} &  \multicolumn{3}{|c|}{Detection Rate} & \multicolumn{3}{|c|}{False Alarm Rate}\\  \hline \hline
256 & 8 & 100.00 & 100.00 & 100.00 & 100.00 & 9.00 & 6.80 \\ 
  256 & 10 & 100.00 & 100.00 & 100.00 & 100.00 & 9.00  & 6.80 \\ 
  256 & 13 & 100.00 & 100.00 & 100.00 & 100.00 & 9.00  & 6.80 \\ 
  256 & 15 & 100.00 & 100.00 & 100.00 & 100.00 & 9.00  & 6.80 \\ 
   \hline
512 & 5 & 100.00 & 36.00 & 25.20 & 99.80 & 5.00 & 1.65 \\ 
  512 & 10 & 100.00 & 100.00 & 100.00 & 99.80 & 5.00 & 1.65 \\ 
  512 & 15 & 100.00 & 100.00 & 100.00 & 99.80 & 5.00 & 1.65 \\ 
  512 & 20 & 100.00 & 100.00 & 100.00 & 99.80 & 5.00 & 1.65 \\ 
   \hline
\end{tabular}
\end{adjustbox}
   \end{table*}

\section{Applying anomaly detection algorithms to count networks}

Often the main interest for monitoring social networks is to study the change of the communication level among the entities in a network or sub-network. This communication level can be represented as the number of communications between two entities $i$ and $j$ which is usually modeled by a Poisson distribution or some variant \citep{woodall2017overview}. We refer to these networks as count networks. Social network communications and transportation networks are some examples of count networks. 

In this section, we study the behavior and performance of the spectral methods proposed by \cite{miller2015spectral} in count networks and evaluate their performance for detecting anomalies. For binary networks, these anomalies had the form of cliques or bipartite subgraphs. In count networks, anomalies can be presented as a small sub-networks with a greater propensity to connect with respect to the rest of the network. By studying the level of communications between the different entities, we can identify unexpected relationships among some of the nodes of the network.
 
For count networks, $A_{ij}$ is the number of edges between vertices $i$ and $j$, for $i,j=1,\dots, n$ and we assume $A_{ij} \sim Poisson(\lambda_{ij})$ under some network model. As in the binary case, the algorithms applied here use the spectral structure of the residual matrix $B=A-E[A]$ to detect any anomaly in the network. In \cite{miller2015spectral},  the Erd\"os-R\'enyi (ER) model, the R-MAT model, and the Chung-Lu model were used. The results presented in this section are based on the ER and Chung-Lu models, since the R-MAT model has only been applied to binary networks.

\subsection{Network models}

As seen in the binary case, the networks under the Erd\"os-R\'enyi model are generated by a single parameter. In count networks, to generate the entries of the adjacency matrix we use $\lambda_{ij}=\lambda_0$ $\forall i,j$, and hence 
$E[A]$ is $\lambda_0*\textbf{1} \cdot \textbf{1}'$. 

The Chung-Lu model is a more realistic random graph model that has been proven to describe well  the behavior of social networks \citep{AieChungLu2001spectral}. This model is specified by a degree sequence that satisfies a power law. In count networks, to generate the expected degree sequence $k_1,k_2,\dots k_n$ of the background graph, we assume $k_i \sim Pareto(\eta, \theta)$  for $i=1,... n$, where $\eta$ and $\theta$ are the location and shape parameters of the Pareto distribution respectively.  
Under the Chung-Lu model, we use $\lambda_{ij}= c k_ik_j$ where $c$ is a constant, and $E[A_{ij}]=\lambda_{ij}$.


\subsection{Evaluating statistical properties of the algorithms in count networks when there is no anomaly}

When conducting an anomaly detection method, we need to investigate the three criteria that were outlined in Section 3.2. In this section we investigate the first two criteria for count networks: When there is no anomaly, a) the statistic should follow the benchmark distribution, and b) false alarm rates should stay close to target values. Through simulations, we first evaluate the behavior of the statistic by comparing the empirical distribution of the statistic and the theoretical distribution. We do this visually by using histograms and Q-Q plots. Then, we study if the upper quantiles for the empirical distribution of the statistic are close to the upper quantiles of the theoretical (or benchmark) distribution. 

We analyzed different scenarios considering network order of $n=128, 256, 512, 1024$ and different network connectivity. For the ER model, we considered $\lambda_0=0.2, 1, 3$, and for the Chung-Lu model we used different values of the location parameter $\eta=0.133,0.333,1$ and one value of the shape parameter $\theta=1.2$ for the Pareto distribution. We report here results from just a few of those scenarios. The others are part of the supplementary material and the observations are similar to those shown here. 

\subsection{Statistical properties of the chi-square algorithm}
\begin{figure*}[ht!]
	\begin{center}
		\includegraphics[height=85mm,keepaspectratio=true]{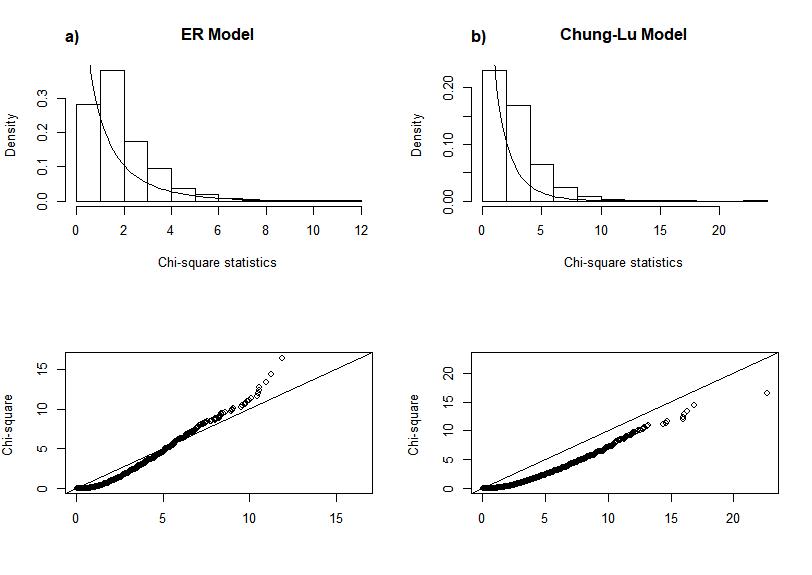}
		\caption{Top figures are histogram density plots of the chi-square statistics based on 10,000 simulations with chi-square distribution, $df = 1$, overlaid. $n = 256$ and $\lambda_0=1$. Bottom figures are the corresponding Q-Q plots. ((a) Erd\"os-R\'enyi, and (b) Chung-Lu Models) }
		\label{chiSQ_plot}
	\end{center}
\end{figure*}

Similar to the binary networks, we compared the empirical distribution of the chi-square statistic from count networks with the theoretical $\chi^2_1$ distribution. Figure \ref{chiSQ_plot} and Table \ref{chisq-1} show the results for some of the explored scenarios.

For the ER model, we observe that the simulated chi-square statistic quantiles are close to the theoretical quantiles of the chi-square distribution with $df=1$. However, in general it is clear that this statistic does not follow the chi-square distribution. Table \ref{chisq-1} shows that the empirical chi-square quantiles can be much higher than the theoretical $\chi^2_1$ quantiles. It is evident that the chi-square algorithm depends on both the network order and the background connectivity. For each scenario, the Kolmogorov-Smirnov (KS) test was performed to see how well the empirical values follow a $\chi^2_1$ distribution. In all cases, we rejected the hypothesis that they follow such distribution. Based on these results, we do not recommend its use for anomaly detection in count networks.

\begin{table*}[ht!]
	\normalsize
	\caption{ Chi-square statistic quantiles from the simulation results compared to chi-square with $df=1$ theoretical quantiles for count networks. (ER and Chung-Lu Models)}
	\label{chisq-1}
	\begin{adjustbox}{width=1\textwidth}
		\begin{tabular}{|rl|lllll|l|lllll|}
			\hline
			\multicolumn{2}{|l|} {} &  \multicolumn{5}{|c|}{ER Model} &  \multicolumn{6}{|c|}{Chung-Lu Model} \\  \hline  
			Network order & $\lambda_0$ & $95\%$ & $96\%$ & $97\%$ & $98\%$ & $99\%$& $\eta$ &$95\%$ & $96\%$ & $97\%$ & $98\%$ & $99\%$\\ 
			\hline \hline
			
			128 & 0.2 &  5.358 & 5.761 & 6.142 & 6.809 & 8.000 & 0.133 & 11.927 & 12.589 & 13.783 & 15.033 & 17.384 \\ 
			& 1 & 4.470 & 4.588 & 5.010 & 5.375 & 6.186 & 0.333 & 7.108 & 7.580 & 8.071 & 9.013 & 10.376 \\ 
			& 3 & 4.174 & 4.476 & 4.639 & 5.217 & 5.971 & 1 & 10.012 & 10.329 & 11.306 & 12.549 & 14.620 \\
			\hline 
			256 & 0.2 &  4.898 & 5.128 & 5.595 & 6.123 & 6.918 &  0.133 & 15.067 & 16.209 & 17.579 & 19.465 & 22.894 \\ 
			& 1 & 4.302 & 4.573 & 5.049 & 5.549 & 6.366 & 0.333 & 8.223 & 8.881 & 9.669 & 10.614 & 12.253\\ 
			& 3 & 4.277 & 4.523 & 4.898 & 5.374 & 6.115  & 1 & 9.080 & 9.756 & 10.548 & 11.436 & 13.250 \\
			\hline 
			512 & 0.2 & 4.557 & 4.885 & 5.270 & 5.696 & 6.539 &   0.133 & 16.201 & 17.252 & 18.787 & 20.584 & 23.602 \\ 
			& 1 & 4.322 & 4.553 & 4.900 & 5.297 & 6.163 & 0.333 & 20.006 & 21.421 & 23.306 & 25.951 & 30.825 \\ 
			& 3 & 4.149 & 4.462 & 4.839 & 5.276 & 6.101 & 1 & 9.069 & 9.615 & 10.563 & 11.668 & 13.785 \\ 
			\hline 
			1024 & 0.2 & 4.443 & 4.651 & 5.037 & 5.566 & 6.448 & 0.133 & 15.482 & 16.509 & 17.774 & 20.160 & 23.781 \\ 
			& 1 & 4.284 & 4.543 & 4.858 & 5.324 & 6.193 & 0.333& 45.341 & 48.821 & 53.603 & 60.493 & 70.091 \\ 
			& 3 & 4.142 & 4.371 & 4.731 & 5.135 & 5.892 & 1& 39.852 & 42.707 & 46.341 & 51.126 & 60.596 \\ 
			\hline 
			& \textbf{$\chi^2_1$ quantiles} & \textbf{3.841} & \textbf{4.218} & \textbf{4.709} &\textbf{ 5.412} & \textbf{6.635}& &\textbf{3.841} & \textbf{4.218} & \textbf{4.709} &\textbf{ 5.412} & \textbf{6.635}\\ 
			\hline
		\end{tabular}
	\end{adjustbox}
\end{table*}

\subsection{Statistical properties of the $L_1$ norm algorithm}

According to \cite{miller2015spectral}, when there is no anomaly present, the $L_1$ norm follows a Gumbel distribution with $a_m$ and $b_m$ as the location and scaling parameter respectively. Neither how to estimate these parameters nor the effect of $m$, the number of eigenvectors, on the performance of the statistics is discussed in \cite{miller2015spectral}. For count networks, we studied also two techniques to estimate the Gumbel distribution parameters, the Method of Moments (MOM) as shown in equation \eqref{eq_am_MOM} with 1000 networks as historical data, and the Extreme Value Theorem as shown in Equation \eqref{eq_am}.

As seen in the binary case,  we studied the effect of $m$ on the performance of the $L_1$ norm statistic when there is no anomaly by analyzing two scenarios: $m<n$ and $m=n$. We expect better performance under the $m=n$ scenario since the statistic will contain more information about the network.

\subsubsection*{Estimating $a_m$ and $b_m$ using historical data}

\begin{figure*}[!ht]
	\begin{center}
		\includegraphics[height=85mm,keepaspectratio=true]{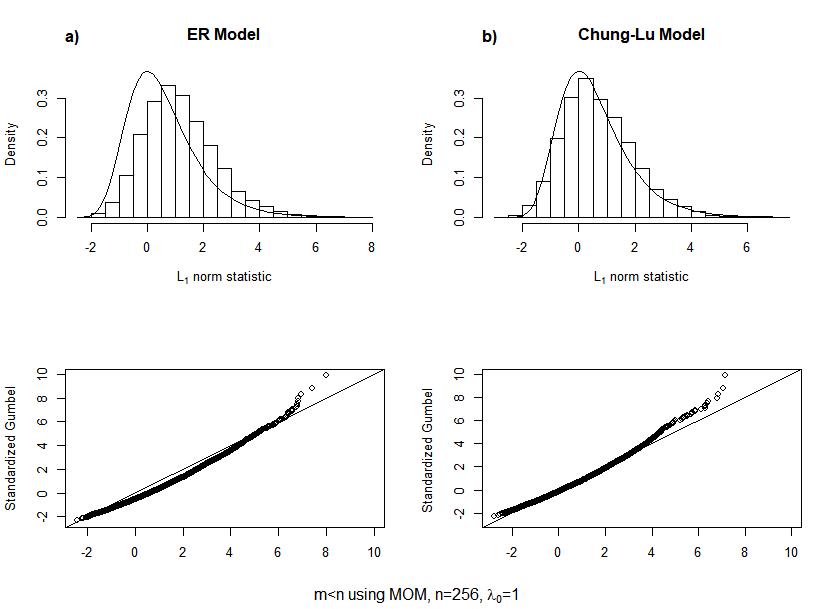}
		\caption{Top figures are histogram density plots of the $L_1$ norm statistics based on 10,000 simulations using MOM estimation and $m<n$, with Gumbel distribution overlaid. $n = 256$ and $\lambda_0=1$. Bottom figures are the corresponding Q-Q plots.((a) Erd\"os-R\'enyi, and (b) Chung-Lu Models)}
		\label{L1_plot_mEV}
	\end{center}
\end{figure*}

As outlined in Section 4, to calculate $L$, first, we  estimate $\mu_k$ and $\sigma_k$ for $k=1,\dots m$. We generated 1000 random networks to obtain such estimates. Then, the $L_1$ norm statistics were calculated for both cases, $m<n$ and $m=n$. For the $m<n$ case, the results from the simulations are shown in Figure \ref{L1_plot_mEV} and Table \ref{L1-mEV-1}. For the ER model, the empirical quantiles from the simulation are lower than the theoretical Gumbel quantiles. For the Chung-Lu model, the empirical quantiles have a high variance, and are different from the theoretical Gumbel quantiles in most cases.

\begin{figure}[!ht]
	\begin{center}
		\includegraphics[height=85mm,keepaspectratio=true]{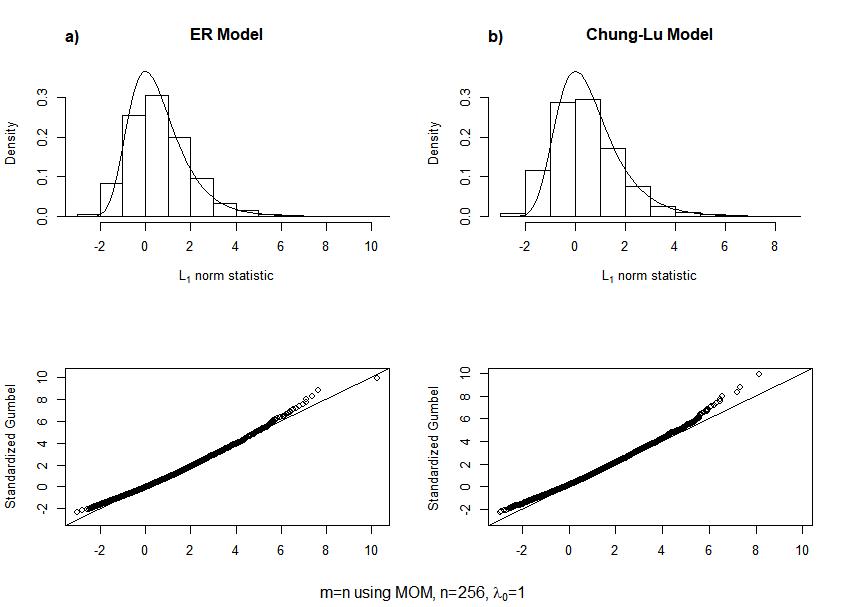}
		\caption{Top figures are histogram density plots of the $L_1$ norm statistics based on 10,000 simulations using MOM estimation and $m=n$, with Gumbel distribution overlaid. $n = 256$ and $\lambda_0=1$. Bottom figures are the corresponding Q-Q plots.((a) Erd\"os-R\'enyi, and (b) Chung-Lu Models) }
		\label{L1_plot_allEV}
	\end{center}
\end{figure}

Figure \ref{L1_plot_allEV} and Table \ref{L1-allEV-1} show the results when using MOM estimators and $m=n$. As expected, for the $m=n$ case, this algorithm performs better. The empirical $L_1$ norm quantiles from the simulations are closer to the theoretical Gumbel quantiles than those of the $m<n$ case. However, the empirical quantiles are frequently higher than the theoretical ones.

\subsubsection*{Estimating $a_m$ and $b_m$ using the Extreme Value Theorem}

\begin{figure*}[ht!]
	\begin{center}
		\includegraphics[height=85mm,keepaspectratio=true]{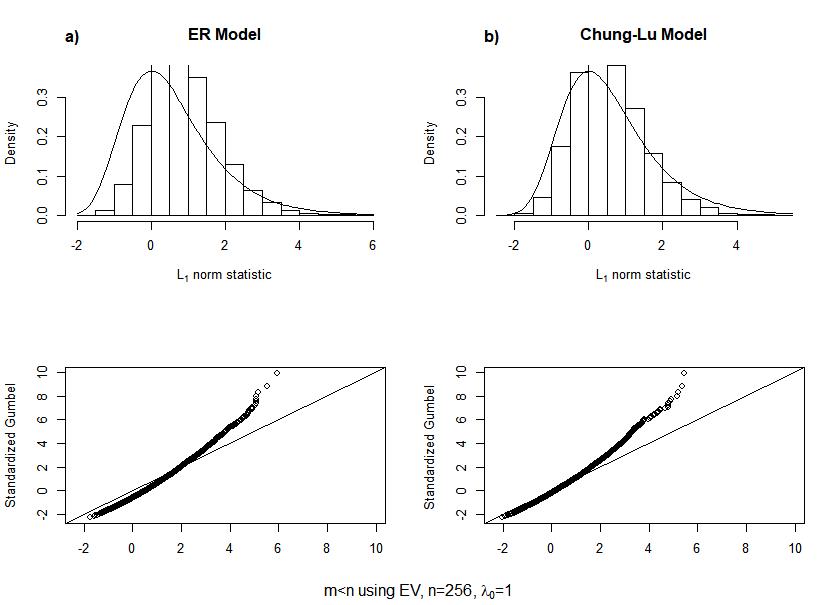}
		\caption{Top figures are histogram density plots of the $L_1$ norm statistics based on 10,000 simulations using Extreme Value Theorem and $m<n$, with Gumbel distribution overlaid. $n=256$ and $\lambda_0=1$. Bottom figures are the corresponding Q-Q plots. ((a) Erd\"os-R\'enyi, and (b) Chung-Lu Models)}
		\label{L1_plot_mEV1}
	\end{center}
\end{figure*}

We also used Extreme Value Theorem to estimate the parameters for the Gumbel distribution, and considered two cases: $m<n$ and $m=n$, as before. For the $m<n$ case, some simulation cases are shown in Figure \ref{L1_plot_mEV1} and Table \ref{L1-mEV-2}. These results of our simulation confirm that the $L_1$ norm algorithm performs better in count networks when using the Extreme Value Theorem. Although the empirical quantiles are not that close to the theoretical Gumbel quantiles, they are less variable than the detection statistic results when using MOM estimation. We observe that the quantiles are independent of the network order and graph connectivity.

\begin{table*}[!ht]
	\caption{$L_1$ norm percentiles when \textbf{$m<n$} using MOM estimation from the simulation results compared to the Gumbel theoretical quantiles for count networks. (ER Model and Chung-Lu Models)}
	\label{L1-mEV-1}
	\begin{adjustbox}{width=1\textwidth}
		\begin{tabular}{|rl|lllll|l|lllll|}
			\hline
			\multicolumn{2}{|l|} {} &  \multicolumn{5}{|c|}{ER Model} &  \multicolumn{6}{|c|}{Chung-Lu Model} \\  \hline  
			Network order & $\lambda_0$ & $95\%$ & $96\%$ & $97\%$ & $98\%$ & $99\%$& $\eta$ &$95\%$ & $96\%$ & $97\%$ & $98\%$ & $99\%$\\ 
			\hline  \hline
			
			128 & 0.2 & 1.623 & 1.837 & 2.111 & 2.483 & 3.140 & 0.133 & 3.548 & 3.883 & 4.341 & 4.943 & 6.012 \\ 
			& 1 & 1.259 & 1.474 & 1.727 & 2.137 & 2.701 & 0.333 & -0.361 & -0.122 & 0.172 & 0.533 & 1.120 \\ 
			& 3 & 1.233 & 1.423 & 1.737 & 2.113 & 2.732  & 1 & 1.126 & 1.329 & 1.625 & 1.961 & 2.529 \\ 
			\hline
			256 & 0.2 & 1.360 & 1.620 & 1.916 & 2.311 & 2.949 & 0.133 & 1.743 & 1.966 & 2.219 & 2.614 & 3.176 \\ 
			& 1 & 1.181 & 1.411 & 1.681 & 2.046 & 2.601 & 0.333 & 0.408 & 0.591 & 0.846 & 1.157 & 1.738 \\ 
			& 3 & 1.119 & 1.307 & 1.554 & 1.900 & 2.500 & 1 & 0.418 & 0.620 & 0.851 & 1.268 & 1.960 \\ 
			\hline
			512 & 0.2 & 1.216 & 1.426 & 1.704 & 2.069 & 2.689 & 0.133 & 1.089 & 1.298 & 1.574 & 1.983 & 2.557 \\ 
			& 1 & 1.006 & 1.233 & 1.535 & 1.931 & 2.486 & 0.333 & 1.664 & 1.869 & 2.088 & 2.353 & 2.801 \\ 
			& 3 & 1.082 & 1.244 & 1.510 & 1.850 & 2.365 & 1 & -0.908 & -0.705 & -0.453 & -0.102 & 0.491 \\ 
			\hline
			1024 & 0.2 & 1.011 & 1.220 & 1.486 & 1.843 & 2.514  & 0.133 & 3.345 & 3.547 & 3.812 & 4.207 & 4.770 \\ 
			& 1 & 0.956 & 1.137 & 1.368 & 1.672 & 2.310 & 0.333 & 3.087 & 3.673 & 4.493 & 5.368 & 6.952 \\ 
			& 3 & 0.904 & 1.064 & 1.317 & 1.682 & 2.239 & 1 & -2.041 & -1.865 & -1.638 & -1.352 & -0.777 \\ 
			\hline
			&		\textbf{Gumbel quantiles} & \textbf{2.970} & \textbf{3.199} & \textbf{3.491} & \textbf{3.902} & \textbf{4.600} & & \textbf{2.970} & \textbf{3.199} & \textbf{3.491} & \textbf{3.902} & \textbf{4.600}\\ 
			\hline
		\end{tabular}
	\end{adjustbox}
\end{table*}

\begin{figure*}[ht!]
	\begin{center}
		\includegraphics[height=80mm,keepaspectratio=true]{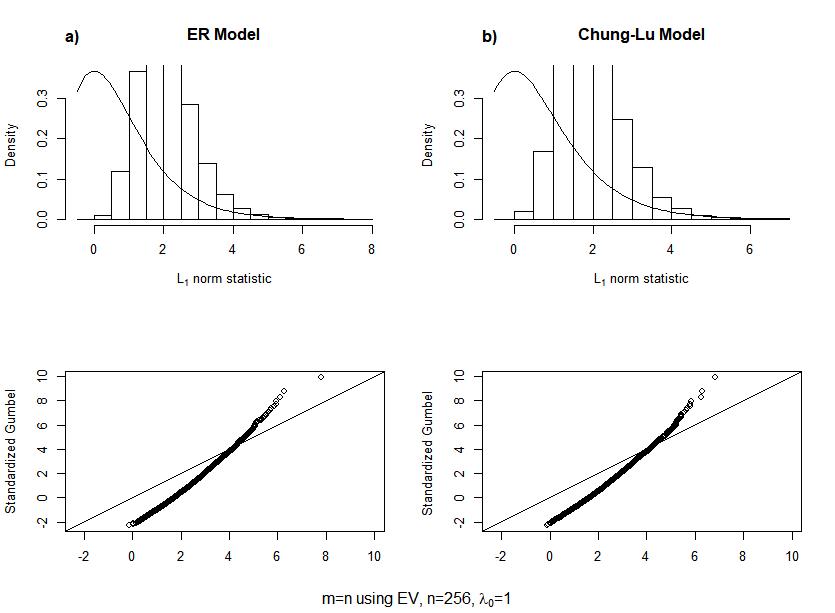}
		\caption{Top figures are histogram density plots of the $L_1$ norm statistics based on 10,000 simulations using Extreme Value Theorem and $m=n$, with Gumbel distribution overlaid. $n=256$ and $\lambda_0=1$. Bottom figures are the Q-Q plots of the simulation. ((a) Erd\"os-R\'enyi, and (b) Chung-Lu Models) }
		\label{L1_plot_allEV2}
		\end{center}
		\end{figure*}

Finally, Figure \ref{L1_plot_allEV2} and Table \ref{L1-allEV-2} present the result when $m=n$, showing that the $L_1$ norm empirical quantiles are closer to the Gumbel theoretical quantiles than to those from the other cases analyzed. We observe that the performance of the $L_1$ norm when using Extreme Value Theorem and $m=n$ does not depend on the order of the network or the graph connectivity. 

In general for count networks, we observe that when using Extreme Value Theorem to estimate the parameters for the Gumbel distribution the $L_1$ norm statistics is closer to Gumbel distribution under both graph models. This conclusion in important since the algorithms evaluated in this papers are intended for static networks for which no historical network data are available.

\begin{table*}[!ht]
	\caption{$L_1$ norm percentiles when $m=n$ using MOM estimation from the simulation results compared to the Gumbel theoretical quantiles for count networks.(ER and Chung-Lu Models)}
	\label{L1-allEV-1}
	\begin{adjustbox}{width=1\textwidth}
		\begin{tabular}{|rl|lllll|l|lllll|}
			\hline
			\multicolumn{2}{|l|} {} &  \multicolumn{5}{|c|}{ER Model} &  \multicolumn{6}{|c|}{Chung-Lu Model} \\  \hline  
			Network order & $\lambda_0$ & $95\%$ & $96\%$ & $97\%$ & $98\%$ & $99\%$& $\eta$ &$95\%$ & $96\%$ & $97\%$ & $98\%$ & $99\%$\\ 
			\hline  \hline
			
			128 & 0.2 & 4.240 & 4.436 & 4.689 & 5.006 & 5.642 & 0.133 & 6.437 & 6.799 & 7.274 & 7.938 & 9.109 \\  
			& 1 & 4.057 & 4.237 & 4.469 & 4.703 & 5.208 & 0.333 & 2.745 & 2.935 & 3.179 & 3.534 & 4.114 \\ 
			& 3 & 3.941 & 4.117 & 4.338 & 4.651 & 5.129 & 1 & 3.903 & 4.064 & 4.300 & 4.546 & 5.085 \\ 
			
			\hline
			256 & 0.2 & 4.465 & 4.634 & 4.886 & 5.217 & 5.778 & 0.133 & 5.434 & 5.638 & 5.897 & 6.258 & 6.896 \\  
			& 1 & 4.332 & 4.522 & 4.734 & 5.063 & 5.506 & 0.333 & 3.980 & 4.144 & 4.353 & 4.686 & 5.169 \\ 
			& 3 & 4.315 & 4.487 & 4.710 & 5.024 & 5.452 & 1 & 4.105 & 4.303 & 4.511 & 4.874 & 5.363 \\ 
			\hline
			512 & 0.2 & 4.783 & 4.937 & 5.147 & 5.410 & 5.877 & 0.133 & 4.794 & 4.984 & 5.178 & 5.460 & 5.963 \\ 
			& 1 & 4.681 & 4.849 & 5.087 & 5.350 & 5.839 & 0.333 & 5.271 & 5.438 & 5.630 & 5.866 & 6.284 \\  
			& 3 & 4.730 & 4.878 & 5.059 & 5.355 & 5.789  & 1 & 4.207 & 4.376 & 4.581 & 4.829 & 5.332 \\ 
			\hline
			1024 & 0.2 & 5.013 & 5.171 & 5.362 & 5.660 & 6.044 & 0.133 & 6.619 & 6.783 & 6.947 & 7.184 & 7.652 \\  
			& 1 &  4.975 & 5.113 & 5.291 & 5.540 & 6.034  & 0.333 & 6.030 & 6.562 & 7.143 & 8.050 & 9.338 \\ 
			& 3 & 4.981 & 5.136 & 5.300 & 5.518 & 5.985 & 1 & 4.466 & 4.632 & 4.889 & 5.149 & 5.504 \\ 
			\hline
			&		\textbf{Gumbel quantiles} & \textbf{2.970} & \textbf{3.199} & \textbf{3.491} & \textbf{3.902} & \textbf{4.600} & & \textbf{2.970} & \textbf{3.199} & \textbf{3.491} & \textbf{3.902} & \textbf{4.600}\\ 
			\hline
		\end{tabular}
	\end{adjustbox}
\end{table*}


\begin{table*}[ht!]
	\caption{$L_1$ norm percentiles when \textbf{$m<n$} using Extreme Value Theorem from the simulation results compared to the Gumbel theoretical quantiles for count networks. (ER and Chung-Lu Models) }
	\label{L1-mEV-2}
	\begin{adjustbox}{width=1\textwidth}
		\begin{tabular}{|rl|lllll|l|lllll|}
			\hline
			\multicolumn{2}{|l|} {} &  \multicolumn{5}{|c|}{ER Model} &  \multicolumn{6}{|c|}{Chung-Lu Model} \\  \hline  
			Network order & $\lambda_0$ & $95\%$ & $96\%$ & $97\%$ & $98\%$ & $99\%$& $\eta$ &$95\%$ & $96\%$ & $97\%$ & $98\%$ & $99\%$\\ 
			\hline  \hline
			
			128 & 0.2 & 2.934 & 3.078 & 3.262 & 3.512 & 3.955 & 0.133 & 4.329 & 4.562 & 4.880 & 5.298 & 6.041 \\  
			& 1 & 2.688 & 2.833 & 3.003 & 3.280 & 3.659  & 0.333 & 1.612 & 1.778 & 1.983 & 2.234 & 2.641 \\  
			& 3 & 2.671 & 2.799 & 3.010 & 3.264 & 3.680 & 1 & 2.646 & 2.787 & 2.992 & 3.226 & 3.621 \\ 
			\hline
			256 & 0.2 & 2.756 & 2.932 & 3.131 & 3.397 & 3.827 & 0.133 & 3.074 & 3.229 & 3.405 & 3.680 & 4.070 \\
			& 1 & 2.636 & 2.791 & 2.973 & 3.219 & 3.592 & 0.333 & 2.147 & 2.274 & 2.451 & 2.667 & 3.071 \\ 
			& 3 & 2.594 & 2.721 & 2.887 & 3.120 & 3.524 & 1 & 2.154 & 2.294 & 2.455 & 2.744 & 3.225 \\  
			\hline
			512 & 0.2 & 2.527 & 2.685 & 2.895 & 3.170 & 3.638 & 0.133 & 2.483 & 2.645 & 2.860 & 3.178 & 3.625 \\ 
			& 1 & 2.368 & 2.540 & 2.767 & 3.066 & 3.484  & 0.333 & 2.930 & 3.090 & 3.260 & 3.466 & 3.815 \\ 
			& 3 & 2.426 & 2.548 & 2.749 & 3.005 & 3.393 & 1 & 0.928 & 1.086 & 1.283 & 1.556 & 2.018 \\ 
			\hline
			1024 & 0.2 & 2.372 & 2.530 & 2.731 & 3.000 & 3.506 & 0.133 & 4.238 & 4.395 & 4.601 & 4.909 & 5.347 \\ 
			& 1 & 2.331 & 2.467 & 2.641 & 2.871 & 3.352 & 0.333 & 4.037 & 4.494 & 5.131 & 5.812 & 7.045 \\ 
			& 3 & 2.291 & 2.412 & 2.603 & 2.878 & 3.298 & 1 & 0.047 & 0.184 & 0.361 & 0.583 & 1.031 \\
			\hline
			&		\textbf{Gumbel quantiles} & \textbf{2.970} & \textbf{3.199} & \textbf{3.491} & \textbf{3.902} & \textbf{4.600} & & \textbf{2.970} & \textbf{3.199} & \textbf{3.491} & \textbf{3.902} & \textbf{4.600}\\ 
			\hline
		\end{tabular}
	\end{adjustbox}
\end{table*}

\begin{table*}[!ht]
	\caption{$L_1$ norm percentiles when $m=n$ using Extreme Value Theorem from the simulation results compared to the Gumbel theoretical quantiles estimates for count networks. (ER Model and Chung-Lu models) }
	\label{L1-allEV-2}
	\begin{adjustbox}{width=1\textwidth}
		\begin{tabular}{|rl|lllll|l|lllll|}
			\hline
			\multicolumn{2}{|l|} {} &  \multicolumn{5}{|c|}{ER Model} &  \multicolumn{6}{|c|}{Chung-Lu Model} \\  \hline  
			Network order & $\lambda_0$ & $95\%$ & $96\%$ & $97\%$ & $98\%$ & $99\%$& $\eta$ &$95\%$ & $96\%$ & $97\%$ & $98\%$ & $99\%$\\ 
			\hline  \hline
			
			128 & 0.2 & 3.157 & 3.353 & 3.604 & 3.922 & 4.556 & 0.133 & 5.280 & 5.638 & 6.108 & 6.766 & 7.925 \\ 
			& 1 & 2.974 & 3.154 & 3.385 & 3.619 & 4.123 & 0.333 & 1.624 & 1.813 & 2.054 & 2.405 & 2.980 \\ 
			& 3 & 2.858 & 3.034 & 3.255 & 3.567 & 4.044 & 1 & 2.771 & 2.931 & 3.164 & 3.407 & 3.941 \\ 
			\hline
			256 & 0.2 & 3.076 & 3.262 & 3.537 & 3.901 & 4.517  & 0.133 & 4.071 & 4.294 & 4.575 & 4.969 & 5.664 \\ 
			& 1 & 2.930 & 3.138 & 3.371 & 3.733 & 4.219  & 0.333 & 2.488 & 2.666 & 2.894 & 3.256 & 3.783 \\ 
			& 3 & 2.911 & 3.100 & 3.344 & 3.690 & 4.159 & 1 & 2.624 & 2.840 & 3.066 & 3.462 & 3.994 \\ 
			\hline
			512 & 0.2 & 3.064 & 3.248 & 3.497 & 3.811 & 4.368 & 0.133 & 3.010 & 3.235 & 3.463 & 3.797 & 4.391 \\ 
			& 1 & 2.943 & 3.143 & 3.426 & 3.740 & 4.322 & 0.333 & 3.573 & 3.770 & 3.997 & 4.276 & 4.770 \\  
			& 3 & 3.001 & 3.178 & 3.394 & 3.745 & 4.263  & 1 & 2.316 & 2.516 & 2.758 & 3.051 & 3.645 \\ 
			\hline
			1024 & 0.2 & 2.928 & 3.129 & 3.374 & 3.755 & 4.246 & 0.133 & 4.890 & 5.098 & 5.305 & 5.606 & 6.199 \\ 
			& 1 & 2.879 & 3.055 & 3.283 & 3.601 & 4.234  & 0.333 & 4.143 & 4.817 & 5.554 & 6.705 & 8.338 \\
			& 3 & 2.886 & 3.084 & 3.295 & 3.573 & 4.171 & 1 & 2.159 & 2.369 & 2.695 & 3.026 & 3.475 \\ 
			\hline
			&		\textbf{Gumbel quantiles} & \textbf{2.970} & \textbf{3.199} & \textbf{3.491} & \textbf{3.902} & \textbf{4.600} & & \textbf{2.970} & \textbf{3.199} & \textbf{3.491} & \textbf{3.902} & \textbf{4.600}\\ 
			\hline
		\end{tabular}
	\end{adjustbox}
\end{table*}

\subsection{Evaluating the performance of the chi-square and $L_1$ norm algorithms in count networks}
As described in Section 4, we used false alarm rates and detection rates to evaluate the performance of the algorithms. Different scenarios were considered based on the network order $n=128, 256,512$, and $1024$, and background connectivity of $\lambda_0=0.2,1$, and $3$ for the ER model and $\eta=0.133,0.333$, and $1$ for the Chung-Lu model. We ran 500 simulations for each combination of network order and connectivity, 250 of which had an anomaly and 250 did not have an anomaly. 

In count networks, an anomaly is presented as a small network with a greater propensity to connect. For each combination, we randomly embedded anomalous subgraphs of 2\%, 5\%, 7\% and 10\% of the network order. To generate the anomaly we used  $\lambda_1=\lambda_0+\delta$ for the ER model, and  $\lambda_{1ij}=\lambda_{0ij}+\delta$ for the Chung-Lu model. The values $\delta=0.5, 2,$ and $3$ were used.

Since our results from the previous section suggest that the $L_1$ norm statistics using MOM estimation does not work well, in this section we studied the performance of the chi-square statistics and $L_1$ norm statistics using Extreme Value Theorem. We compared these statistics obtained from our simulations to a threshold $K$ obtained from the benchmark distribution. If the statistic exceeds the threshold, an anomaly is signaled. 

We considered different values of $K$. For the chi-square algorithm, the quantile of the chi-square distribution with $df=1$ at $\alpha=0.05, 0.02,$ and $0.01$ were studied as thresholds. For the $L_1$ norm algorithm, we used the quantiles of the Gumbel distribution at $\alpha=0.05, 0.02,$ and $0.01$. Here we present the results when $\alpha=0.05$ since the results from the other cases showed similar patterns.

\begin{figure*}[ht!] 
	\begin{center}
		\includegraphics[height=80mm,keepaspectratio=true]{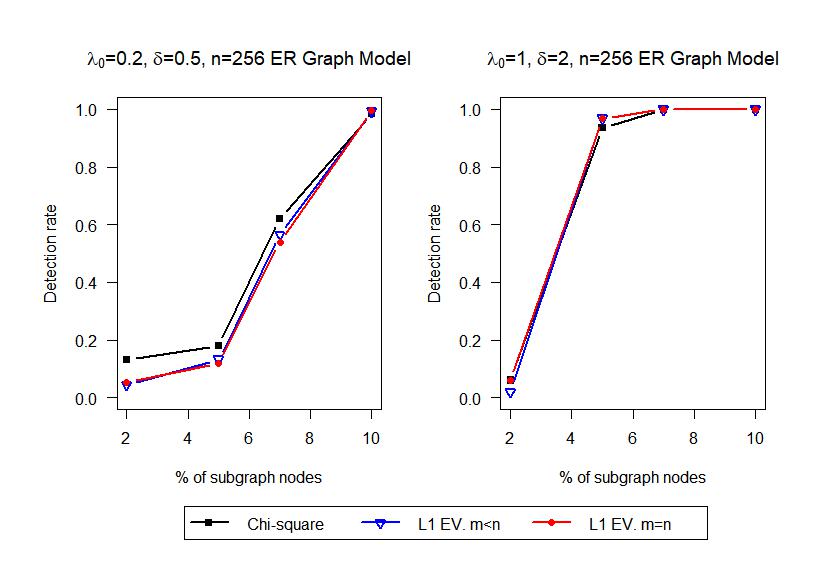}
		\caption{Detection rates for count networks with $n = 256$. Percentage of anomalous subgraph nodes varies from 2\%, 5\%, 7\%, and 10\% of $n = 256$. (Erd\"os-R\'enyi Model) }
		\label{detRate95_2}
	\end{center}
\end{figure*}

\begin{figure*}[ht!]
	\begin{center}
		\includegraphics[height=80mm,keepaspectratio=true]{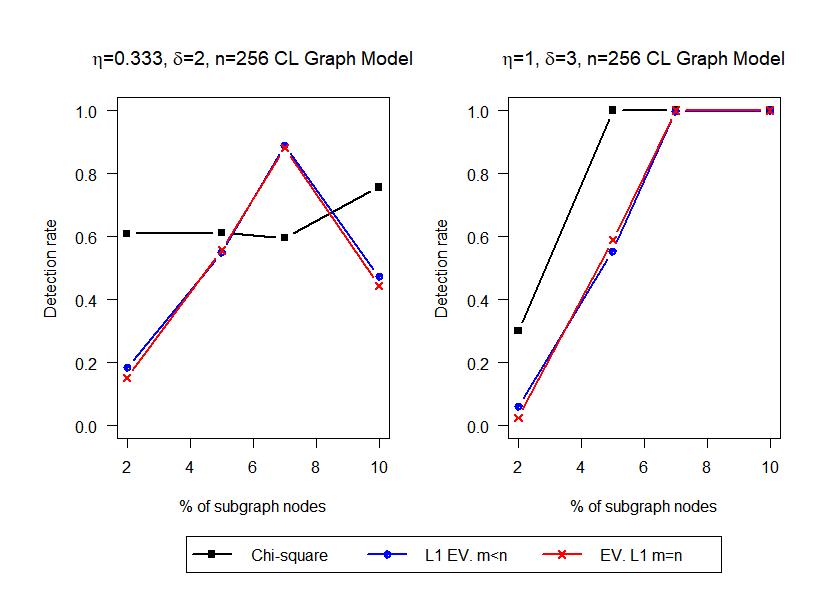}
		\caption{Detection rates for count networks with $n = 256$. Percentage of anomalous subgraph nodes varies (2\%, 5\%, 7\%, and 10\%
			of $n = 256$). (Chung-Lu Model) }
		\label{detRate95_4}
	\end{center}
\end{figure*}

Table \ref{falseAlarms} shows the false alarm rate results for the different scenarios. The chi-square algorithm, in all scenarios explored, gave a higher false alarm rate than the desired 0.05, especially for the Chung-Lu model, while the $L_1$ norm algorithm performs much better. Figures \ref{detRate95_2} and \ref{detRate95_4} show the detection rates for different graph connectivity values under the ER and Chung-Lu models respectively. We observe that the $L_1$ norm algorithm performs similarly in both cases, $m=n$ and $m<n$ . None of the detection algorithms perform well when the proportion of anomalous nodes is small (2\%). 

For the ER model, the three algorithms perform better as the number of anomalous nodes increases.  For the the Chung-Lu model, the $L_1$ norm algorithm resulted in a non-monotone behavior with respect to the percentage of anomalous nodes present in count networks. For most of the scenarios analyzed and shown in the supplementary material, the detection rate of the $L_1$ algorithm increases as the number of anomalous nodes increases up to some point, then decreases to then improve again. 

This non-monotone behavior of the $L_1$ norm was also observed in \cite{miller2015spectral}. 
When embedding a number of anomalous nodes in the background network, the eigenvalues in the residual matrix form two clusters due to a model mismatch. 
In \cite{miller2015spectral}, the authors reported a similar phenomenon of eigenvalue clustering that leads to this non-monotonic behavior of detection rates.
This leads to this non-monotone behavior in the performance of the $L_1$ algorithm. 
This phenomenon was also observed in some binary networks that are shown in the supplementary material.

In general, we observe that the use of the chi-square method does not work well in count networks. Our results suggest that the $L_1$ norm algorithm using Extreme Value Theorem when $m<n$ is a better choice for detecting anomalies in count networks although it presents some problems as mentioned above.

\begin{table*}[ht!]
	\centering
	\caption{Comparison of false alarm rates for the three cases, $\chi^2$ algorithm, $L_1$ norm using Extreme Value Theorem and $m<n$, and $L_1$ norm using Extreme Value Theorem and $m=n$. $95^{th}$ percentile was used for signaling threshold.}
	\label{falseAlarms}
	\begin{adjustbox}{width=1\textwidth}
		\begin{tabular}{|r|ccc|ccc|ccc|}
			\hline
			Network order && chi-square&&& $L_1$ norm $m<n$ &&& $L_1$ norm $m=n$ &\\
			\hline
			ER model & $\lambda_0=0.2$ & $\lambda_0=1$ & $\lambda_0=3$ & $\lambda_0=0.2$ & $\lambda_0=1$ & $\lambda_0=3$ & $\lambda_0=0.2$ & $\lambda_0=1$ & $\lambda_0=3$\\ 
			\hline
			128 & 0.11 & 0.06 & 0.08 & 0.04 & 0.02 & 0.04 & 0.06 & 0.04 & 0.05 \\ 
			256 & 0.10 & 0.06 & 0.07 & 0.04 & 0.02 & 0.01 & 0.08 & 0.07 & 0.04 \\ 
			512 & 0.08 & 0.05 &  0.08  & 0.01 &  0.02 & 0.00 & 0.06 & 0.04 & 0.04 \\ 
			1024 & 0.07 & 0.07 & 0.11 & 0.01 & 0.01 & 0.01 & 0.04 & 0.06 & 0.04 \\  
			\hline
			Chung-Lu model & $\eta_0=0.133$ & $\eta_0=0.333$ & $\eta_0=1$ & $\eta_0=0.133$ & $\eta_0=0.333$ & $\eta_0=1$ & $\eta=0.133$ & $\eta_0=0.333$ & $\eta_0=1$\\ 
			\hline
			128 & 0.36 & 0.35 & 0.47 & 0.00 & 0.04 & 0.01 & 0.09 & 0.06 & 0.04 \\ 
			256 & 0.55 & 0.63 & 0.28 & 0.03 & 0.16 & 0.01 & 0.09 & 0.16 & 0.02 \\ 
			512 & 0.45 & 0.46 & 0.65 & 0.01 & 0.16 & 0.00 & 0.03 & 0.19 & 0.00 \\ 
			1024 & 0.66 & 0.70 & 0.53 & 0.01 & 0.01 & 0.08 & 0.03 & 0.01 & 0.25 \\  
			\hline
		\end{tabular}
	\end{adjustbox}
\end{table*}

\section{Discussion and Future Work}
In our paper, we evaluated two spectral algorithms proposed in \cite{miller2015spectral} for anomaly detection in binary, static networks.  It is implicitly  assumed in \cite{miller2015spectral} that the chi-square detection statistic follows the $\chi^2$ distribution, while the $L_1$ norm detection statistic follows the Gumbel distribution. We show that this is not the case by comparing the quantiles obtained from multiple simulation studies where we compared the detection statistic values to their respective theoretical distributions. Specifically, we show that the distributions of these values are affected by the connectivity of the network, the network order, and the network model. 

These inconsistencies, such as the different behaviors when applied to different network orders, network models, and connectivities, means the algorithms are impractical to a practitioner. For example, it is difficult to establish a signaling detection threshold due to these inconsistencies. We also show that because the $L_1$ norm algorithm requires historical data for implementation, it is unsuitable for a practitioner to use for static networks.  

Many of these concerns are addressed in our paper. We introduced improvements to the chi-square algorithm that improved its performance in sparse networks. The resulting detection and false alarm rates after our improvements show that our recommendations are advantageous over the current algorithm. We also proposed a way of standardizing the $L_1$ norm statistic that requires only the currently observed network. We compared the effects of our improvements to the theoretical distributions and show that they perform sufficiently well. 

Finally, we extended these algorithms to count networks, an area of importance to practitioners, but something not investigated in \cite{miller2015spectral}. The algorithms along with our improvements perform sufficiently well when applied to count networks and the same conclusions were obtained. Future research involves further extending these algorithms to dynamic networks. 
Statistical evaluation of anomaly detection methods is an important research area where little work has been done, and we encourage more work in this important direction.

\section{Acknowledgments}
We would like to thank the Area Editor and two anonymous Reviewers for their constructive and helpful suggestions. 

\subsection*{Conflicts of interest}
The authors have no conflicts of interest to disclose.

\renewcommand\refname{Bibliography}

\bibliographystyle{apalike} 
\bibliography{spectral} 

\begin{thebibliography}{}

\bibitem[Aiello et~al., 2001]{AieChungLu2001spectral}
Aiello, W., Chung, F., and Lu, L. (2001).
\newblock A random graph model for power law graphs.
\newblock {\em Experimental Mathematics}, 10(1):53--66.

\bibitem[Akoglu et~al., 2010]{akoglu2010oddball}
Akoglu, L., McGlohon, M., and Faloutsos, C. (2010).
\newblock Oddball: Spotting anomalies in weighted graphs.
\newblock {\em Advances in Knowledge Discovery and Data Mining}, pages
  410--421.

\bibitem[Akoglu et~al., 2015]{akoglu2015graph}
Akoglu, L., Tong, H., and Koutra, D. (2015).
\newblock Graph based anomaly detection and description: a survey.
\newblock {\em Data Mining and Knowledge Discovery}, 29(3):626--688.

\bibitem[Albert et~al., 2004]{albert2004structural}
Albert, R., Albert, I., and Nakarado, G.~L. (2004).
\newblock Structural vulnerability of the {N}orth {A}merican power grid.
\newblock {\em Physical review E}, 69(2):025103.

\bibitem[Azarnoush et~al., 2016]{azarnoush2016monitoring}
Azarnoush, B., Paynabar, K., Bekki, J., and Runger, G. (2016).
\newblock Monitoring temporal homogeneity in attributed network streams.
\newblock {\em Journal of Quality Technology}, 48(1):28--43.

\bibitem[Bader and Madduri, 2008]{bader2008snap}
Bader, D.~A. and Madduri, K. (2008).
\newblock Snap, small-world network analysis and partitioning: An open-source
  parallel graph framework for the exploration of large-scale networks.
\newblock In {\em IEEE International Symposium on Parallel and Distributed
  Processing, 2008}, pages 1--12. IEEE.

\bibitem[Cer et~al., 2012]{cer2012searching}
Cer, R., Bruce, K., Donohue, D., Temiz, N., Mudunuri, U., Yi, M., Volfovsky,
  N., Bacolla, A., Luke, B., Collins, J., and Stephens, R. (2012).
\newblock Searching for non-{B} {DNA}-forming motifs using n{BMST} (non-{B DNA}
  motif search tool).
\newblock {\em Current protocols in human genetics}, pages 18.7.1--18.7.22.

\bibitem[Cer et~al., 2011]{cer2011introducing}
Cer, R.~Z., Bruce, K.~H., Donohue, D.~E., Temiz, A.~N., Bacolla, A., Mudunuri,
  U.~S., Yi, M., Volfovsky, N., Luke, B.~T., and Collins, J.~R. (2011).
\newblock Introducing the non-{B} {DNA} {M}otif {S}earch {T}ool (n{BMST}).
\newblock {\em Genome biology}, 12(1):P34.

\bibitem[Chakrabarti et~al., 2004]{chakrabarti2004r}
Chakrabarti, D., Zhan, Y., and Faloutsos, C. (2004).
\newblock R-{MAT}: A recursive model for graph mining.
\newblock In {\em Proceedings of the 2004 SIAM International Conference on Data
  Mining}, pages 442--446. SIAM.

\bibitem[Chawla and Sun, 2006]{Chawla2006}
Chawla, S. and Sun, P. (2006).
\newblock {SLOM}: a new measure for local spatial outliers.
\newblock {\em Knowledge and Information Systems}, 9(4):412--429.

\bibitem[Chung et~al., 2003]{chung2003spectra}
Chung, F., Lu, L., and Vu, V. (2003).
\newblock Spectra of random graphs with given expected degrees.
\newblock {\em Proceedings of the National Academy of Sciences},
  100(11):6313--6318.

\bibitem[Dahan et~al., 2017]{dahan2017network}
Dahan, M., Sela, L., and Amin, S. (2017).
\newblock Network monitoring under strategic disruptions.
\newblock {\em arXiv preprint arXiv:1705.00349}.

\bibitem[Erdos and R{\'e}nyi, 1960]{erdos1960evolution}
Erdos, P. and R{\'e}nyi, A. (1960).
\newblock On the evolution of random graphs.
\newblock {\em Publ. Math. Inst. Hung. Acad. Sci}, 5(1):17--60.

\bibitem[Farahani et~al., 2017]{mazrae2016statistical}
Farahani, E.~M., Kazemzadeh, R.~B., Noorossana, R., and Rahimian, G. (2017).
\newblock A statistical approach to social network monitoring.
\newblock {\em Communications in Statistics-Theory and Methods},
  46(22):11272--11288.

\bibitem[Haveliwala, 2003]{haveliwala2003topic}
Haveliwala, T.~H. (2003).
\newblock Topic-sensitive pagerank: A context-sensitive ranking algorithm for
  web search.
\newblock {\em IEEE Transactions on Knowledge and Data Engineering},
  15(4):784--796.

\bibitem[Lei and Rinaldo, 2015]{lei2015consistency}
Lei, J. and Rinaldo, A. (2015).
\newblock Consistency of spectral clustering in stochastic block models.
\newblock {\em The Annals of Statistics}, 43(1):215--237.

\bibitem[Mall et~al., 2013]{mall2013kernel}
Mall, R., Langone, R., and Suykens, J.~A. (2013).
\newblock Kernel spectral clustering for big data networks.
\newblock {\em Entropy}, 15(5):1567--1586.

\bibitem[Miller et~al., 2010a]{miller2010L1norm}
Miller, B., Bliss, N., and Wolfe, P.~J. (2010a).
\newblock Subgraph detection using eigenvector {L}1 norms.
\newblock {\em Advances in Neural Information Processing Systems},
  23:1633--1641.

\bibitem[Miller et~al., 2015]{miller2015spectral}
Miller, B.~A., Beard, M.~S., Wolfe, P.~J., and Bliss, N.~T. (2015).
\newblock A spectral framework for anomalous subgraph detection.
\newblock {\em IEEE Transactions on Signal Processing}, 63(16):4191--4206.

\bibitem[Miller et~al., 2010b]{miller2010toward}
Miller, B.~A., Bliss, N.~T., and Wolfe, P.~J. (2010b).
\newblock Toward signal processing theory for graphs and non-{E}uclidean data.
\newblock In {\em Proceedings of the Acoustics Speech and Signal Processing
  (2010)}, pages 5414--5417. ICASSP.

\bibitem[Nadarajah and Kotz, 2004]{nadarajah2004beta}
Nadarajah, S. and Kotz, S. (2004).
\newblock The beta {G}umbel distribution.
\newblock {\em Mathematical Problems in Engineering}, (4):323--332.

\bibitem[Newman, 2016]{newman2016community}
Newman, M. (2016).
\newblock Community detection in networks: {M}odularity optimization and
  maximum likelihood are equivalent.
\newblock {\em arXiv preprint arXiv:1606.02319}.

\bibitem[Papadimitriou et~al., 2003]{papadimitriou2003loci}
Papadimitriou, S., Kitagawa, H., Gibbons, P.~B., and Faloutsos, C. (2003).
\newblock Loci: Fast outlier detection using the local correlation integral.
\newblock In {\em Proceedings of the 19th International Conference on Data
  Engineering (2003)}, pages 315--326.

\bibitem[Priebe et~al., 2005]{priebe2005scan}
Priebe, C.~E., Conroy, J.~M., Marchette, D.~J., and Park, Y. (2005).
\newblock Scan statistics on {E}nron graphs.
\newblock {\em Computational \& Mathematical Organization Theory},
  11(3):229--247.

\bibitem[Procter et~al., 2010]{procter2010visualization}
Procter, J.~B., Thompson, J., Letunic, I., Creevey, C., Jossinet, F., and
  Barton, G.~J. (2010).
\newblock Visualization of multiple alignments, phylogenies and gene family
  evolution.
\newblock {\em Nature methods}, 7:S16--S25.

\bibitem[Qin and Rohe, 2013]{qin2013regularized}
Qin, T. and Rohe, K. (2013).
\newblock Regularized spectral clustering under the degree-corrected stochastic
  blockmodel.
\newblock In {\em Advances in Neural Information Processing Systems}, pages
  3120--3128.

\bibitem[Ranshous et~al., 2015]{FaloustousDynamic}
Ranshous, S., Shen, S., Koutra, D., Harenberg, S., Faloutsos, C., and Samatova,
  N.~F. (2015).
\newblock Anomaly detection in dynamic networks: a survey.
\newblock {\em Wiley Interdisciplinary Reviews: Computational Statistics},
  7(3):223--247.

\bibitem[Raulf-Heimsoth et~al., 1998]{raulf1998analysis}
Raulf-Heimsoth, M., Chen, Z., Rihs, H., Kalbacher, H., Liebers, V., and Baur,
  X. (1998).
\newblock Analysis of t-cell reactive regions and {HLA-DR4} binding motifs on
  the latex allergen {H}ev b 1 (rubber elongation factor).
\newblock {\em Clinical and Experimental Allergy}, 28(3):339--348.

\bibitem[Rohe et~al., 2011]{rohe2011spectral}
Rohe, K., Chatterjee, S., and Yu, B. (2011).
\newblock Spectral clustering and the high-dimensional stochastic blockmodel.
\newblock {\em The Annals of Statistics}, 39(4):1878--1915.

\bibitem[{\v{S}}altenis, 2004]{vsaltenis2004outlier}
{\v{S}}altenis, V. (2004).
\newblock Outlier detection based on the distribution of distances between data
  points.
\newblock {\em Informatica}, 15(3):399--410.

\bibitem[Savage et~al., 2014]{savage2014anomaly}
Savage, D., Zhang, X., Yu, X., Chou, P., and Wang, Q. (2014).
\newblock Anomaly detection in online social networks.
\newblock {\em Social Networks}, 39:62--70.

\bibitem[Sengupta, 2018]{sengupta2018anomaly}
Sengupta, S. (2018).
\newblock Anomaly detection in static networks using egonets.
\newblock {\em arXiv preprint arXiv:1807.08925}.

\bibitem[Sengupta and Chen, 2015]{sengupta2015spectral}
Sengupta, S. and Chen, Y. (2015).
\newblock Spectral clustering in heterogeneous networks.
\newblock {\em Statistica Sinica}, 25:1081--1106.

\bibitem[Singh et~al., 2011]{MillerSparsePCA}
Singh, N., Miller, B.~A., Bliss, N.~T., and Wolfe, P.~J. (2011).
\newblock Anomalous subgraph detection via sparse principal component analysis.
\newblock In {\em 2011 IEEE Statistical Signal Processing Workshop (SSP)},
  pages 485--488.

\bibitem[Sun et~al., 2005]{sun2005neighborhood}
Sun, J., Qu, H., Chakrabarti, D., and Faloutsos, C. (2005).
\newblock Neighborhood formation and anomaly detection in bipartite graphs.
\newblock In {\em Fifth IEEE International Conference on Data Mining (2005)},
  pages 1--8.

\bibitem[Wang et~al., 2012]{wang2012identify}
Wang, G., Xie, S., Liu, B., and Yu, P.~S. (2012).
\newblock Identify online store review spammers via social review graph.
\newblock {\em ACM Transactions on Intelligent Systems and Technology (TIST)},
  3(4):61.

\bibitem[Woodall et~al., 2017]{woodall2017overview}
Woodall, W.~H., Zhao, M.~J., Paynabar, K., Sparks, R., and Wilson, J.~D.
  (2017).
\newblock An overview and perspective on social network monitoring.
\newblock {\em IISE Transactions}, 49(3):354--365.

\end{thebibliography}

\end{document}